\documentclass[aps,prb,reprint,groupedaddress]{revtex4-1}
\usepackage[dvipdfmx]{graphicx} 
\usepackage{color}

\usepackage{amsmath,amssymb} 
\usepackage{amsmath} 
\usepackage{amsfonts} 
\usepackage{bm} 
\newcommand{\cket}[1]{| #1 \rangle}

\usepackage{ulem}

\newcommand{\e}{{\rm e}}
\newcommand{\im}{{\rm i}}

\begin{document}
\preprint{HEP/123-qed}
\title{Sine-square deformation applied to classical Ising models}
\author{Chisa Hotta}
\affiliation{Graduate School of Arts and Sciences, The University of Tokyo, 3-8-1, Komaba, Meguro-ku, Tokyo 1538902, Japan. }

\author{Takashi Nakamaniwa}
\affiliation{Graduate School of Arts and Sciences, The University of Tokyo, 3-8-1, Komaba, Meguro-ku, Tokyo 1538902, Japan. }

\author{Tota Nakamura}
\affiliation{College of Engineering, Shibaura Institute of Technology, Saitama 337-8570, Japan. }

\date{\today}
\begin{abstract}
Sine-square deformation (SSD) is a treatment proposed in quantum systems, 
which spatially modifies a Hamiltonian, 
gradually decreasing the local energy scale from the center of the system toward the edges by a sine-squared envelope function. 
It is known to serve as a good boundary condition as well as 
to provide physical quantities reproducing those of the infinite-size systems. 
We apply the SSD to one- and two-dimensional classical Ising models. 
Based on the analytical calculations and Monte Carlo simulations, 
we find that the classical SSD system is regarded as an extended canonical ensemble of 
a local subsystem each characterized by its own effective temperature. 
This effective temperature is defined by normalizing the system temperature by the deformed local energy scale. 
A single calculation for a fixed system temperature provides a set of physical quantities 
of various temperatures that quantitatively reproduces well those of the uniform system. 
\end{abstract}
\pacs{02.70.Rr,05.50.+q, 75.10.Hk,64.60.Cn}
\maketitle
\narrowtext 
\section{Introduction}
\begin{figure}[t]
  \begin{center}
   \includegraphics[width=85mm]{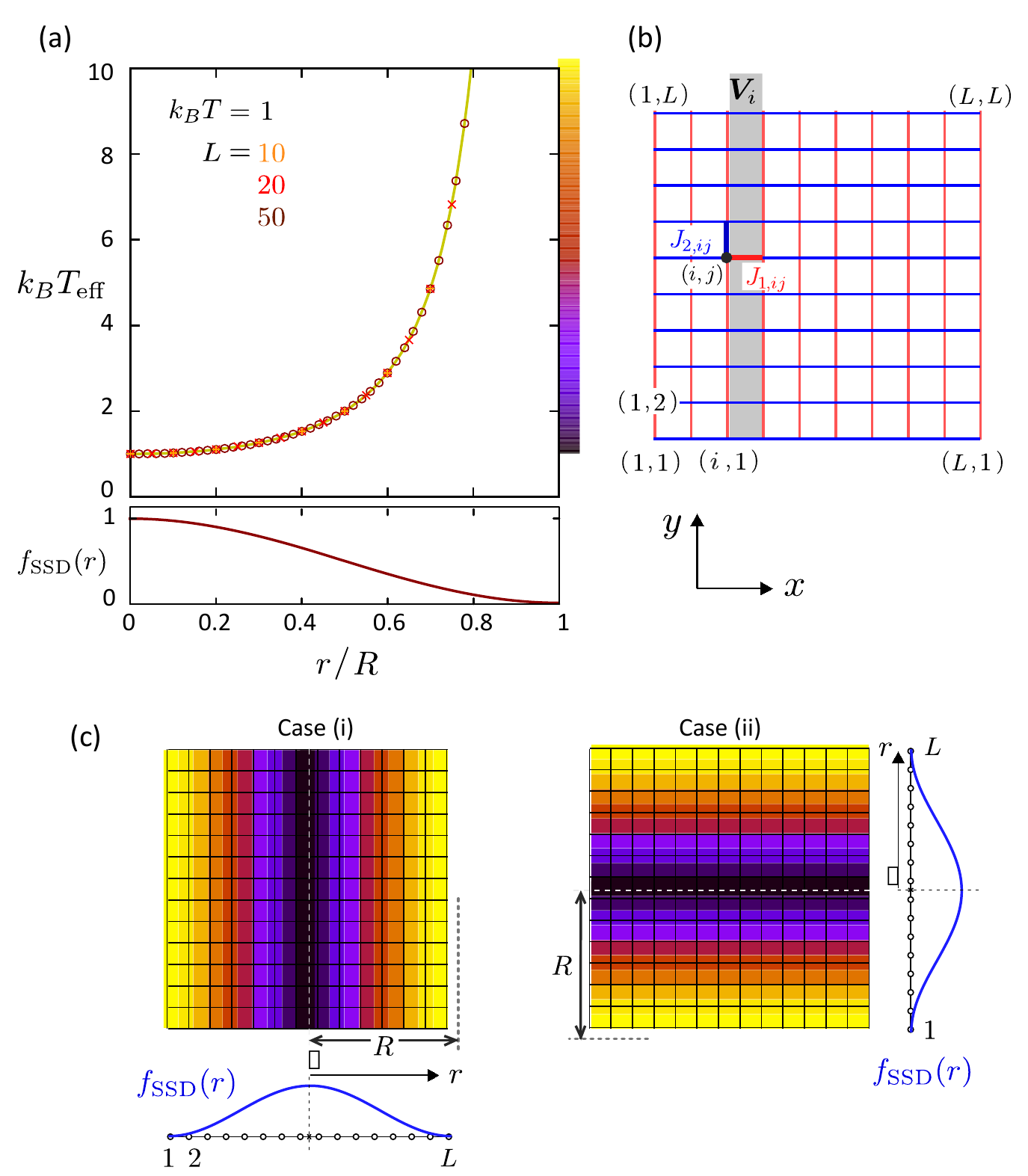}
	  \caption{ (a) Change in effective temperature, $k_BT_{\rm eff}$, from the center ($r=0$) toward the edge ($r/R=1$), 
    and the corresponding $f_{\rm SSD}(r)$.
  (b) Two-dimensional square lattice with interactions $J_{1,ij}, J_{2,ij}$ 
      on bonds running in the positive directions of $x$- and $y$-axes 
      from site $\gamma=(i,j)$. 
  The transfer matrix $\bm V_i$ is constructed in a unit of $L$ sites along the $y$-direction. 
  (c) SSD along row (Case (i)) and column (Case (ii)). }
    \label{f1}
  \end{center} 
\end{figure}

The Hamiltonian in condensed matter physics is spatially uniform in most cases, 
and its symmetry determines the physical properties of the system. 
For this reason, deforming the Hamiltonian may usually mean modifying the physical state itself. 
However, this turned out not always to be the case 
for a series of operation called ``sine-square deformation'' (SSD). 
The deformed Hamiltonian, ${\cal H}_{\rm SSD}$, 
is generated from the original Hamiltonian, 
${\cal H}=\int d\bm r {\cal H}(\bm r)$, using an envelope function, $f_{\rm SSD}(\bm r)$, as 
\begin{equation}
	{\cal H}_{\rm SSD}= \int d \bm r f_{\rm SSD} (\bm r){\cal H}(\bm r), 
\end{equation}
where $\bm r$ is a coordinate of the system with its origin at the center, and 
\begin{equation}
f_{\rm SSD} (\bm r)= \frac{1}{2}\Big(1+\cos \big(\frac{\pi r}{R}\big)\Big), 
\label{fssd}
\end{equation} 
with $R$ being half the system length. 
As shown in Fig.\ref{f1}, the sine-square function, $f_{\rm SSD}$, governs 
the whole system by a single wavelength of $2R$, 
and $f_{\rm SSD}(\bm r){\cal H}(\bm r)$ varies smoothly 
from the maximum value at the center toward the edges. 
The SSD was proposed as one of the smooth boundary conditions to remove boundary effects\cite{nishino09,gendiar11}, 
e.g. Friedel oscillations from the open boundaries, or 
artificial potentials 
that emerges for a chosen cluster sizes and shapes which may stabilize fictitious orders\cite{shibata11}. 
The SSD Hamiltonian is empirically known to generate the wave function that recovers the translational symmetry, 
in perfect coincidence with the wave function under the periodic boundary condition (PBC) \cite{gendiar11,hikihara11}. 
This coincidence is proved in an XY model and a transverse-field Ising chain\cite{katsura11,katsura11-2}. 
The SSD ground state is also shown to serve as a restricted class of conformal field theories with some applications to a wider class of conformal mappings\cite{ryu16,okunishi16,katsura17,ishibashi15,kishimoto18,liu20}, and in that context 
the quantum dynamics of the SSD system is tested\cite{wen20,neupert20,ageev21,wen21,fan21}. 
When an external field is applied, 
the physical quantities as a response function to the field is evaluated from the SSD Hamiltonian, 
whose accuracy against the exact solution in the thermodynamic limit is ${\cal O}(10^{-4})$\cite{hotta12}. 
\par
From these studies, it is found that the SSD Hamiltonian loses its translational symmetry 
but could still or better reproduce the physical properties of the original Hamiltonian. A key to understanding this phenomenon is the idea of real-space renormalization\cite{okunishi10}, which stems from Wilson's poor man's scaling\cite{wilson}. 
In a system with translational symmetry, a quantum state is characterized by a wave number. 
When the SSD is imposed, a wavenumber is no longer a quantum number, 
and a scattering term generated by $f_{\rm SSD}(r)$ mixes this original eigen states\cite {maruyama11}. 
Such mixing generates a series of localized wave packets that serves as another basis set of the Hamiltonian. 
Since these states are localized, they are no longer influenced by the system size nor by the boundary. 
Therefore, 
it allows us to obtain physical quantities that reproduce those in the thermodynamic limit\cite{hotta13}. 
\par
In the present paper, we apply this SSD to classical Ising models, as they provide a good platform
to test approximate methods based on their exactly solvable structures\cite{onsager}. 
The Hamiltonian of the {\it uniform} classical Ising model is, 
\begin{equation}
{\cal H}= \sum_{\langle \bm \gamma, \bm \gamma' \rangle} -J \sigma_{\bm \gamma}\sigma_{\bm \gamma'}, 
\label{hising}
\end{equation}
with the uniform coupling constant $J$. The index $\bm \gamma$ 
represents a lattice site and the summation is taken over all the neighboring pairs of spins. 
The exact solutions are known for the one-dimensional(1D) chain and two-dimensional(2D) square lattice. 
A second-order phase transition occurs in the 2D square lattice model at the temperature, 
$k_BT_c = 2/\ln(1+\sqrt{2})\sim 2.2692J$. 
\par
In constructing the SSD Ising Hamiltonian, 
we replace the uniform interaction $J$ with $f_{\rm SSD}(r)J$, 
where the position vector $\bm r$ is defined at the center of each bond. 
Suppose that the temperature of this deformed system is $k_BT$. 
Then, the system can be regarded as an assemblage of Ising spins
with ``uniform" interactions at a renormalized ``effective temperature", 
$k_BT_{\rm eff}\equiv k_BT/f_{\rm SSD}(r)$. 
Figure~\ref{f1}(a) shows an effective temperature as a function of $r$; the minimum value, $k_BT_{\rm eff}(r=0)=k_BT$, 
at the center gradually increases toward infinity at the system edge. 
\par
The deformation in 1D chain is straightforward. 
For the site index, $i=1,\cdots, L$, in Fig.\ref{f1}(c), 
the bond connecting the $i$ and $(i+1)$-th sites is located at $r_i=i-L/2$, and 
we set $R=L/2$ to generate the values $f_{\rm SSD}(r_i)$ in Eq.(\ref{fssd}).

For 2D square lattice, we consider an $L\times L$ lattice shown in Fig.~\ref{f1}(b) 
to keep the aspect ratio as unity\cite{sandvik12}, 
and define a coupling along the row between sites $(i,j)$ and $(i+1,j)$ as $J_{1;ij}$, 
and its location is defined by $\bm r_{1;ij}$. 
We take a bond along the column between sites $(i,j)$ and $(i,j + 1)$ as $J_{2;ij}$ 
located at $\bm r_{2;ij}$. 
Here, the vectors $\bm r_{1;ij}$, $\bm r_{2;ij}$ are not the ordinary position vectors but 
are introduced to define the deformation function 
along the axis $\bm r$ parallel to the row and column for Case (i) and (ii), respectively, 
as shown in Fig.~\ref{f1}(c). 
We consider two cases: 
Case (i) deforms the bond interaction only along the row direction, 
keeping those along the column uniform; we plug in $r_{1;ij}=i-L/2$ and $r_{2;ij}=i-(L-1)/2$ to Eq.(\ref{fssd}) with $R=L/2$. 
The second one, Case (ii) deforms the interaction along the column, while keeping the row direction uniform; 
we take $r_{1;ij}=j-(L-1)/2$ and $r_{2;ij}=j-L/2$ for this case. 
The two cases formally differ in the analytical treatment as we show in \S.\ref{sec_transmat}, 
where we construct the column-to-column transfer matrix. 
\par
The aim of this paper is to clarify the role of SSD in a classical Ising model. 
We show that the energy and related quantities accurately reproduce those of $k_BT_{\rm eff}$, 
which means that one can obtain a set of data 
for a wide temperature range in a single calculation. 
The physical implication is that the SSD system is an assemblage of local subsystems 
with different temperatures, which form a modified canonical ensemble. 
The neighboring local subsystems have similar effective temperatures and work to each other as a heat bath. 
In the final part of the manuscript, we address the possibility of taking other types of deformation. 
\par
The paper is organized as follows; 
In \S.\ref{1dtrmat} we demonstrate that the transfer matrix method is exactly applied to the SSD Ising model in 1D. 
In \S.\ref{sec_transmat}, we analyze the 2D Ising model using a fermionic representation, 
and obtain an exact form of the partition function for a finite system size 
for both Case (i) and Case (ii). 
Then, we numerically evaluate the bond energy 
using these formulations in \S \ref{sec_numerical}. 
We also perform a classical Monte Carlo simulation for the SSD Hamiltonian in \S \ref{sec_mc} 
to test the numerical applicability of SSD. 
\S \ref{sec_final} gives a summary and discussion. 
\par
%
%
\section{Exact solution of one dimensional SSD Ising model}
\noindent
\label{1dtrmat}
Let us first consider a partition function of a 1D periodic lattice consisting of $L$ sites, 
\begin{eqnarray}
Z&=& \sum_{\{\sigma_i\}} {\rm exp} \big({\sum_{i=1}^L K_{i} \sigma_i\sigma_{i+1}}\big)
\nonumber \\
&=& \sum_{\{\sigma_i\}} \prod_{i=1}^L {\rm exp}(K_i\sigma_i\sigma_{i+1})
={\rm Tr}\Big(\prod_{i=1}^L V_i \Big), 
\nonumber \\
V_i&=& \left(\begin{array}{ll}
{\rm e}^{K_i} & {\rm e}^{-K_i} \\
{\rm e}^{-K_i} & {\rm e}^{K_i} 
\end{array}\right)
\label{mat1d}
\end{eqnarray}
using the conventional notation, $K_i \equiv J_i/k_BT$ with $J_i\equiv Jf_{\rm SSD}(r_i)$. 
Here, $r_i=i-L/2$ is the location of bond connecting site $i$ and $i+1$.
The eigenvalues of the transfer matrix $V_i$ are, 
$\lambda_i^{\pm}=\e^{K_i}\pm \e^{-K_i}$, 
which explicitly depend on index-$i$, whereas, 
the corresponding eigen vectors, $\bm p_i^\pm=(1,\pm 1)/\sqrt{2}$, are site-independent. 
In this way, all the transfer matrices are simultaneously diagonalized and 
the partition function is exactly given as 
\begin{equation}
Z=\prod_{i=1}^L \lambda_i^+ + \prod_{i=1}^L \lambda_i^-.
\label{z1d}
\end{equation}
Then, the exact expectation value of the bond energy is obtained as 
\begin{eqnarray}
\langle \sigma_i \sigma_{i+1} \rangle
&=& \frac{(\lambda_i^-/\lambda_i^+)\prod_{l=1}^L \lambda_l^+ + (\lambda_i^+/\lambda_i^-)\prod_{l=1}^L \lambda_l^- }{\prod_{l=1}^L \lambda_l^+ + \prod_{l=1}^L \lambda_l^-}.
\label{bond1d}
\end{eqnarray}
Taking the limit of open boundary, $J_L\rightarrow 0$, 
we find $\lambda_L^+\rightarrow 2$ and $\lambda_L^-\rightarrow 0$, and 
the bond energy $\langle \sigma_i \sigma_{i+1} \rangle $ converges to 
\begin{eqnarray}
\frac{\lambda_i^-}{\lambda_i^+} =\tanh \big(\frac{J_i}{k_BT}\big)
= \tanh \big(\frac{J}{k_BT_{\rm eff}(r_{i})}\big). 
\label{ebulk}
\end{eqnarray}
Reminding that energy per bond in the uniform system takes a form, $e_{\rm bulk} = -\tanh (J/k_BT)$, 
in the thermodynamic limit, 
one finds that the site-dependent bond energy of the SSD Hamiltonian is $e_{\rm bulk}$ 
at their ``local" effective temperatures, $k_BT_{\rm eff}(\bm r_i)$. 
The partition function at $L\rightarrow\infty$ is given as $Z=\prod_i \lambda_i^+$ 
which matches {\it exactly} the partition function of a system consisting of Ising bond degrees of freedom $\lambda_i^+$ 
with interaction $J$ and $i$-dependent effective temperature $k_BT_{\rm eff}(\bm r_i)$. 
\par
For later convenience, we show that Eq.(\ref{mat1d}) is written using the Pauli matrix, 
$\tau^\nu$, $\nu=x,y,z$ (we use $\tau$ instead of $\sigma$ to avoid confusion) 
and unit matrix $I$ and the parameter $K_i^*$ which fulfills $\tanh K_i^* = e^{-2K_i}$ as 
\begin{eqnarray}
V_i &=& \e^{K_i}(I + \tau^x \e^{-2K_i})=(\tanh K_i^*)^{-1/2}( I + \tau^x \tanh K_i^*)
\nonumber\\
&=& (\sinh K_i^* \cosh K_i^*)^{-1/2} (I \cosh K_i^*+ \tau^x \sinh K_i^*)  \nonumber \\
&=& (2\sinh 2K_i)^{1/2} \e^{K_i^*\tau^x}, 
\label{eq:k1}
\end{eqnarray}
where we used a relation $\sinh 2K_i \sinh 2K_i^*=1$. 

\section{Exact solution of two dimensional SSD Ising model}
\label{sec_transmat}
In this section, we expand an analytical formula to compute the partition function $Z$ of the 
SSD Ising model at a finite system size in two dimensions. 
The results shown here are also applied to deformations with functional forms other than $f_{\rm SSD}$. 
Among several different approaches\cite{onsager,kaufman,kac-ward,potts-ward,potts-ward2,kramers-wannier,kramers-wannier2,fisher49,schultz}, 
we build our work based on the analysis 
given by Schultz, Mattis, and Lieb\cite{schultz}. 
It provides a good description of the exact transfer matrix of the 2D Ising model by 
the 1D free fermionic degrees of freedom. 
In \S.\ref{sec:preliminaries}, 
we give a review of this work originally applied to the uniform Hamiltonian and show that it can be applied to our system where the interactions are site-dependent. 
We introduce the operators for the column-to-column transfer matrix in the fermionic representation
and the related trace formula. 
\S.\ref{sec:ssdcases} is devoted to derivations of $Z$ for Cases (i) and (ii) with the SSD Hamiltonian, 
which uses the formula obtained in \S.\ref{sec:preliminaries}. 
\subsection{Preliminaries}
\label{sec:preliminaries}
\subsubsection{Column-to-column transfer matrix }
\label{sec:tr_general}
The explicit form of a transfer matrix, $\bm V_i$, from column-$i$ to column-$(i+1)$ 
is given for a spatially nonuniform Hamiltonian as in the following. 
Let $\bm \sigma_i \equiv\{\sigma_{i,j}| j=1,2,\cdots L \}$ be 
a complete orthonormal basis of dimensions $2^{L}$ of the $i$th column, 
where $\sigma_{i,j}=\pm 1$ is the Ising degrees of freedom on site $(i,j)$ in Fig.\ref{f1}(b). 
The following operators, $\hat V_{1;i}$ and $\hat V_{2;i}$, include the  
interactions along $i$th column ($J_{1;ij}$) and $j$th row ($J_{2;ij}$), respectively, 
and give a partition function $Z$ as 
\begin{eqnarray}
Z &=& {\rm Tr} \big( \prod_{i=1}^L \bm V_{1;i} \bm V_{2;i} \big),\\
\bm V_{1;i} &\equiv&\langle  \bm \sigma_i,\bm \sigma_{i+1} |\hat V_{1;i}|  \bm \sigma_i',\bm \sigma_{i+1}' \rangle \nonumber \\
 &=& \delta_{\bm \sigma_i, \bm \sigma_i'}\delta_{\bm \sigma_{i+1}, \bm \sigma_{i+1}'}
   {\rm exp}\Big( \beta\sum_{j'=1}^L J_{1;ij'} \sigma_{i,j'}\sigma_{i+1,j'}\Big),
\nonumber\\
\bm V_{2;i} &\equiv& \langle \bm \sigma_i|\hat V_{2;i}|  \bm \sigma_i' \rangle 
= \delta_{\bm \sigma_i, \bm \sigma_i'}
{\rm exp}\Big( \beta\sum_{j'=1}^L J_{2;ij'} \sigma_{i,j'}\sigma_{i,j'+1}\Big). 
\nonumber
\label{spintr}
\end{eqnarray}
Generalization of Eq.(\ref{eq:k1}) to $L$ degrees of freedom along the $i$th column immediately gives 
a description of $V_{1;i}$ and $V_{2;i}$ as 
\begin{eqnarray}
\bm V_{1;i}&=& \prod_{j'=1}^L(2\sinh 2K_{1;ij'})^{1/2} 
{\rm exp} \Big( \sum_{j=1}^L  K^*_{1;ij} \bm {\tau^x_{j}}  \Big),
\nonumber\\
\bm V_{2;i}&=& {\rm exp} \Big( \sum_{j=1}^L K_{2;ij}\; \bm {\tau^z_{j}} \bm { \tau^z_{j+1}}  \Big),
\end{eqnarray}
with $K_{1;i,j}=J_{1,ij}/k_BT$ and $K_{2;i,j}=J_{2,ij}/k_BT$, 
and $K^*_{1;ij}$ is defined as $\tanh K_{1;ij}= \e^{-2K_{1;ij}^*}$. 
We use a $2^L\times 2^L$ matrix defined as a direct product, 
$\bm{\tau^\nu_{j}}= I \otimes I \otimes \cdots \otimes \tau^{\nu} \otimes I\cdots \otimes I$, 
with $\tau^{\nu}$ operating on the $j$th Ising degrees of freedom on the column. 
\par
This form is further transformed first 
by rotating the axis of the Pauli matrices as $\tau^x_j \rightarrow -\tau^z_j$ and 
$\tau^z_j \rightarrow \tau^x_j$, 
and then by using a set of Pauli operators $\{\hat \tau^{\nu}_j\}$ operating on $j$th site. 
\begin{eqnarray}
&& \hat V_{1;i}= \big(\prod_{j'=1}^L(2\sinh 2K_{1;ij'})^{1/2}\big) {\rm exp} (\hat H_{1;i}), 
\nonumber\\
&& \hat V_{2;i}= {\rm exp} (\hat H_{2;i}) ,
\nonumber\\
&& \hat H_{1;i}= \sum_{j=1}^L  -K^*_{1;ij} \hat \tau^z_{j}  ,
\nonumber \\
&& \hat H_{2;i}= \sum_{j=1}^L K_{2;ij}\; \hat \tau^x_{j} \hat \tau^x_{j+1}, 
\end{eqnarray}
where the trace of the operators over the $2^N$-Hilbert space gives $Z= {\rm Tr}\big( \prod_{i=1}^L \hat V_{1;i} \hat V_{2;i+1} \big)$. 
\par
By making use of the Jordan-Wigner transformation, 
\begin{eqnarray}
&& \hat \tau^+_j=\frac{1}{2}(\hat \tau^x_j+i \hat \tau^y_j) = {\rm exp} \Big( -\im\pi \sum_{l=1}^{j-1} c_l^\dagger c_l \Big) c_j^\dagger, 
\nonumber\\
&& \hat \tau^-_j=\frac{1}{2}(\hat \tau^x_j-i \hat \tau^y_j) = {\rm exp} \Big( \im\pi \sum_{l=1}^{j-1} c_l^\dagger c_l \Big) c_j, 
\label{jw}
\end{eqnarray}
we obtain a fermionic representation of the operators as 
\begin{eqnarray}
&& \hat H_{1;i} =  -\sum_{j=1}^{L} \!2K_{1;ij}^* \big(c_j^\dagger c_j-\frac{1}{2}\big), 
\label{v1hams}
\\
&& \hat H_{2;i} = \sum_{j=1}^{L} K_{2;ij} \big(c_j^\dagger- c_j\big)\big(c_{j+1}^\dagger + c_{j+1}\big),
\label{v2hams}
\end{eqnarray}
where $c_j^\dagger(c_j)$ is a creation(annihilation) operator of spinless fermion. 
In the standard approach, the partition function is given as 
$Z={\rm Tr}(\prod_{i=1}^L \hat V_i)$ with 
\begin{equation}
\hat V_i=\hat V_{2;i}^{1/2} \hat V_{1;i} \hat V_{2;i+1}^{1/2}
\label{eq:trmatvi}
\end{equation}
being an operator representing the column-$(i)$-to-column-$(i+1)$ transfer matrix. 
If $\hat V_i$ does not depend on $i$, one is able to diagonalize the $2^L\times 2^L$ representation of 
$\hat V_i$, and the product of the largest eigenvalues will give $Z$. 
However, in Case (i), since the transfer matrices depend on column-$i$, 
they cannot be diagonalized simultaneously, and this approach does not straightforwardly apply. 
In the next subsection, we review the derivation of the exact solution for the uniform Ising model. 
The formula Eqs.(\ref{fr}-\ref{eq:h1qh2q}) will be 
adopted to calculate the partition function of Case (i) in the later section.
The eigenvalue solution of the transfer matrix of the uniform system is also to be compared with those obtained for the SSD Hamiltonian. 

\subsubsection{Spatially uniform 2D Ising model}
\label{sec:uniform}
For the interaction parameters, $K_{1;i,j}$ and $K_{2;i,j}$, 
let us omit the row index-$j$ while keep the column index-$i$ to clarify that they do not depend on $j$. 
Since the Hamiltonian is uniform along the column, 
the transfer matrix on the $i$th column is block diagonalized by using the 
Fourier transform of fermionic operators along the $j$-direction, 
\begin{equation}
c_j=\frac{1}{\sqrt{L}} \e^{-\im\pi/4} \sum_q \e^{\im q j} \eta_q, 
\label{fr}
\end{equation}
where we take $q=(2l-1)\pi/L$ and $2\pi l/L$ when the number of fermions, 
${\cal N}=\sum_q \eta_q^\dagger \eta_q$, is even and odd, respectively. 
The even- and odd-${\cal N}$ sectors originate from the constraint imposed on Eq.(\ref{jw}) 
due to anti-periodic (APBC) and periodic(PBC) boundary conditions, respectively. 
The operators in Eqs.(\ref{v1hams}) and (\ref{v2hams}) are rewritten as, 
\begin{eqnarray}
&& \hat H_{1;i}\!=\! \sum_{0\le q <2\pi} \hat H_{1;iq}, 
\rule{50mm}{0mm}\nonumber \\
&& \hat H_{2;i}\!=\! \sum_{0< q <\pi} \hat H_{2;iq} +  (\eta^\dagger_0\eta_0 + \eta^\dagger_\pi\eta_\pi), \nonumber \\
&& \hat H_{1;iq}= -2K_{1;i}^*\big(\eta_q^\dagger \eta_q-\frac{1}{2} \big),
\label{heta_q1}
\\
&& 
\hat H_{2;iq}\!=\!2K_{2;i} \Big(\! \cos q (\eta_q^\dagger \eta_q + \eta_{-q}^\dagger \eta_{-q} ) 
+ \sin q (\eta_q \eta_{-q} + \eta_{-q}^\dagger \eta_{q}^\dagger ) 
\Big),
\nonumber \\
\label{heta_q2}
\end{eqnarray}
where the last two terms of $\hat H_{2;i}$ with wave vectors $q=0,\pi$ 
are present only in the odd-${\cal N}$ sector. 
As $\eta_q$'s with different $|q|$ commute, one can decompose the exponentials of the 
transfer matrix as 
\begin{equation}
\hat V_i= (\sinh 2K_{1;i}^*)^{L/2} \prod_{q} \e^{\hat H_{2;iq}/2}\e^{\hat H_{1;iq}} \e^{\hat H_{2;i+1\,q}/2}. 
\label{eq:vi_q}
\end{equation}
\par
We now prepare a matrix representation of Fock operators 
$\e^{\hat H_{2;iq}/2}$ and $\e^{\hat H_{1;iq}}$ 
by one-body states, $|\bm \eta_{q} \rangle =\bm \eta^\dagger_{q}|0\rangle$, 
where $\bm\eta^\dagger_{q} =(\eta_q^\dagger,\eta_{-q})$. 
\begin{eqnarray}
&&{H_{1;iq}} =\langle \bm \eta_{q} | \hat H_{1;iq} |\bm \eta_{q}\rangle 
=\left(\begin{array}{cc}
-2K_{1;i}^*& 0 \\
0 & 2K_{1;i}^* \!\!
\end{array} \right),\nonumber \\
&& H_{2;iq}= \langle \bm \eta_{q} | \hat H_{2;iq} |\bm \eta_{q}\rangle
=2K_{2;i} \left(\begin{array}{cccc}
\cos q \!& - \sin q \\
-\sin q & -\cos q \!
\end{array} \right). \rule{10mm}{0mm}
\label{eq:h1qh2q}
\end{eqnarray}
\par
The formulation in the rest of this subsection holds only when $H_{2;iq}$ does not depend on 
column index, $i$; 
it is omitted from the interaction parameters as $K_1$ and $K_2$. 
We multiply $\e^{H_{2;iq}/2}=I \cosh K_{2} + (\cos q \tau^z + \sin q \tau^x )\sinh K_{2}$ 
and $\e^{H_{1;iq}}= I\cosh 2K_{1}^* -\tau^z \sinh 2K_{1}^*$ 
using the Pauli matrices $\tau^\nu$, 
and find a real symmetric matrix 
\begin{eqnarray}
&& \e^{\frac{H_{2;iq}}{2}}\e^{H_{1;iq}}\e^{\frac{H_{2;i+1\,q}}{2}}
= \e^{2K_{2}\cos q}\left(\begin{array}{cc}
A_q& C_q \\
C_q & B_q
\nonumber \\
\end{array} \right) 
\label{eq:viuniform}
\nonumber \\
&& = \e^{2K_{2}\cos q} \;P 
\left(\begin{array}{cc}
-2\epsilon^{(u)}_q & 0 \\
0 & 2\epsilon^{(u)}_q 
\end{array} \right) P^{-1}, 
\label{eq:p_2x2}
\end{eqnarray}
where in the final term we diagonalized the matrix by an orthogonal matrix $P$, 
and $\epsilon_q^{(u)}\ge 0$ is obtained by the relationship 
\begin{equation}
\cosh 2\epsilon^{(u)}_q= \cosh2K_{2}\cosh2K_{1}^* -\sinh 2K_2 \sinh2K_1^* \cos q. 
\label{eq:epsilon}
\end{equation}
By the same matrix $P$ as in Eq.(\ref{eq:p_2x2}), the operator $\bm \eta_q$ undergoes 
a Bogoliubov transformation, 
\begin{eqnarray}
&&  \bm \xi_{q} = 
\left(\begin{array}{c}
\xi_{q}\\
\xi^\dagger_{-q}
\end{array}\right) 
= P \bm \eta_q, \;\;
P=\left(\begin{array}{cc}
\cos \phi_{q} & + \sin \phi_{q} \\
- \sin \phi_{q} & \cos \phi_{q} 
\end{array}\right). 
\rule{6mm}{0mm}
\label{eq:bgtr}
\end{eqnarray}
Using this transformation, 
the product part of Eq.(\ref{eq:vi_q}) is rewritten as $\e^{\hat H_i}$, 
where the Fock operator ${\hat H_i}$ representing the transfer matrix is given as
\begin{eqnarray}
\hat V&=&(2\sinh 2K_1)^{N/2} \prod_{i=1}^L \: \e^{\hat H_i}, 
\nonumber \\
\hat H_i &=& \sum_{0<q<\pi} \big(- 2\epsilon^{(u)}_q \xi_q^\dagger \xi_q  + 2\epsilon^{(u)}_q \xi_{-q} \xi_{-q}^\dagger \big)
\nonumber \\
&& -2(K_1^*-K_2)(\eta_0^\dagger\eta_0-\frac{1}{2}) -2(K_1^*+ K_2) (\eta_\pi^\dagger\eta_\pi-\frac{1}{2})
\nonumber \\
&=& -\!\sum_{0\le q<2\pi} 2\epsilon^{(u)}_q(\xi_q^\dagger \xi_q-\frac{1}{2}). 
\label{eq:hq_uniform}
\end{eqnarray}
Here, one can summarize all the $q$-terms by reading off 
$\eta_0=\xi_0$ and $\eta_\pi=\xi_\pi$, since we see from Eq.(\ref{eq:epsilon}) that 
$\epsilon_0=(K_1^*-K_2)$ and $\epsilon_\pi=(K_1^*+K_2)$. 
\par
The partition function is obtained (for reference, see Eq.(\ref{eq:kilch2}) in the next subsection) as
\begin{eqnarray}
Z&=& (2\sinh 2K_1)^{N/2} {\rm Tr}\big(\prod_{i=1}^L \e^{\hat H_i}\big)
\nonumber\\
&=&(2\sinh 2K_1)^{N/2} \e^{L\sum_q\epsilon^{(u)}_q} \prod_q \big(1+\e^{-2L\epsilon^{(u)}_q}\big)
\nonumber\\
&=& (2\sinh 2K_1)^{N/2}\prod_{0\le q <2\pi} 2\cosh(L \epsilon^{(u)}_q). 
\label{eq:zuniform}
\end{eqnarray}
The bond energy is written as 
\begin{eqnarray}
\langle \sigma_{i,j}\sigma_{i,j+1}\rangle
&=&\sum_q -\frac{\tanh(L_1 \epsilon^{(u)}_q)}{\sinh 2\epsilon^{(u)}_q}
\big(\sinh 2K_{2i} \cosh 2K_{1i}^* \nonumber\\
&& \rule{10mm}{0mm} -  \cosh 2K_{2i} \sinh 2K_{1i}^* \cos q )
\label{eq:siguniform1}
\\
\langle \sigma_{i,j}\sigma_{i+1,j}\rangle 
&=&= -\frac{1}{\tanh 2K_{1i}} 
+\sum_q \frac{\tanh(L_1 \epsilon_q^{(u)})  }{\sinh 2\epsilon_q^{(u)} \sinh 2K_{1i}}
\nonumber\\
&& \hspace{-10mm}\times\big(\cosh 2K_{2i} \sinh 2K_{1i}^*  - \sinh 2K_{2i} \cosh 2K_{1i}^* \cos q \big).
\nonumber\\
\label{eq:siguniform2}
\end{eqnarray}
Considering only 
the contributions from the largest term in the product of the second equation in 
Eq.(\ref{eq:zuniform}), which is valid for $L\rightarrow\infty$, we find 
\begin{equation}
Z\rightarrow \Lambda_0 \equiv (2\sinh 2K_1)^{N/2} \e^{L\sum_{0\le q<2\pi} \epsilon_q}, 
\label{eq:z_uniforminfty}
\end{equation}
which reproduces the result in Ref.[\onlinecite{schultz}]. 
This corresponds to the vacuum state of the Bogoliubov quasi-particle, i.e. having $\langle \xi_q^\dagger\xi_q\rangle =0$ for all-$q$. 
In this limit, Eqs.(\ref{eq:siguniform1}) and (\ref{eq:siguniform2}) are modified to those taking 
$\tanh(L_1 \epsilon_q)\rightarrow 1$. 
\par
\begin{figure}[btp]
  \begin{center}
    \includegraphics[width=7cm]{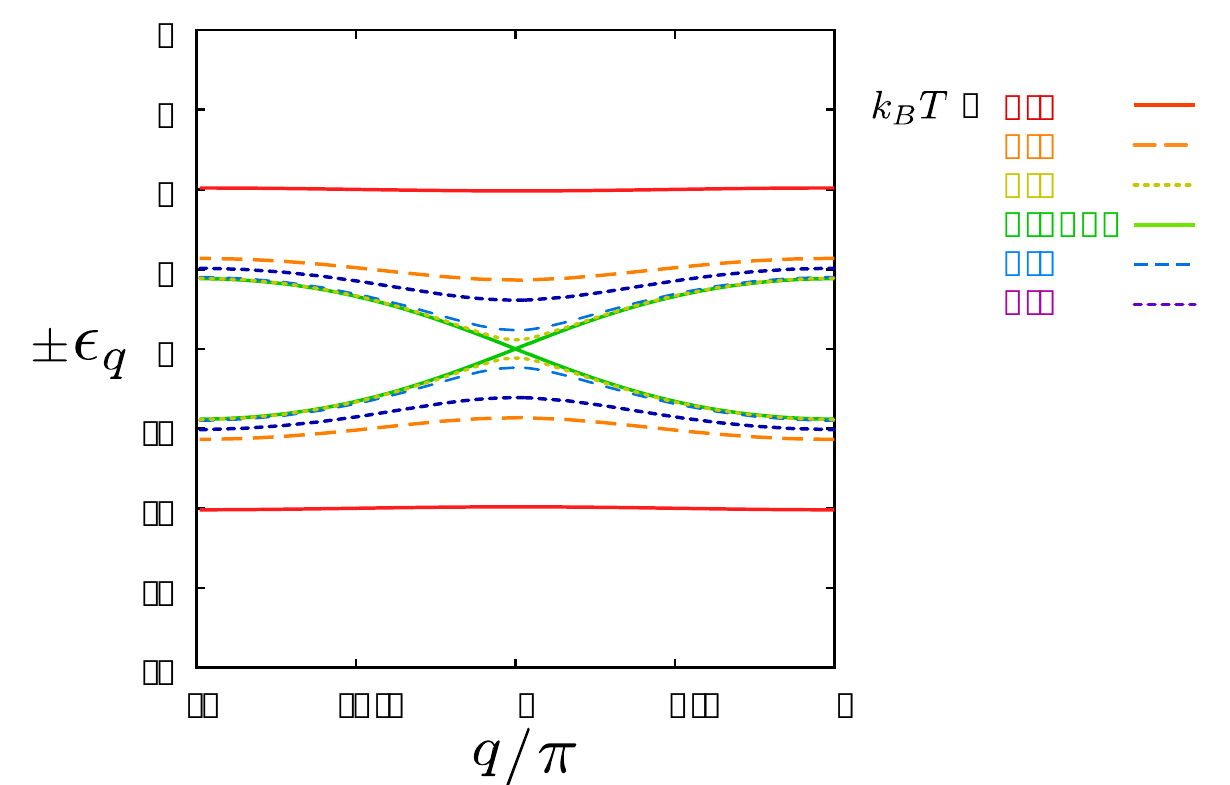}
    \caption{ Dispersion of fermions of a uniform Hamiltonian, $\epsilon_q$, in Eq.(\protect\ref{eq:epsilon}) 
at several different temperatures. 
 }
    \label{f2}
  \end{center}    
\end{figure}

Figure~\ref{f2} shows pairs of $\pm \epsilon^{(u)}_q$ 
at several different temperature. 
Here, by introducing a hole creation operator 
as an anihilation of particle, $\bar \xi_q^\dagger = \xi_q$, 
Eq.(\ref{eq:hq_uniform}) can be rewritten as 
\begin{equation}
\hat H^i= -\!\sum_{0\le q<2\pi} \epsilon^{(u)}_q (\xi_q^\dagger \xi_q- \bar\xi_q^\dagger\bar \xi_q), 
\label{eq:phham}
\end{equation}
where the lowest energy eigen state of $\hat H^i$ is obtained by 
fully occupying a hole band and by keeping particle bands empty. 
This is a ``vacuum"state of Bogoliubov quasi-particle. 
Exciting a Bogoliubov quasi-particle $\langle \xi_q^\dagger\xi_q\rangle \ne 0$ corresponds to 
creating a particle-hole pair with the excitation energy $2\epsilon^{(u)}_q$ at wave number $q$. 
The gap at $q=0$ closes at the transition temperature of the uniform 2D Ising model, $T_c=2.2692$. 
%
%
%
%
%
\subsubsection{Full counting statistics and the trace formula}
\label{sec_klich}
To count the full statistics $Z$ consisting of column-dependent transfer matrices, 
we introduce another basic formula proved by Klich in Ref.[\onlinecite{klich}]. 
Consider a second quantized single-particle operator acting on the Fock space as 
\begin{equation}
\hat \Gamma(\hat A) = \bm c^\dagger A \bm c, \;\; \bm c^\dagger=(c_1^\dagger,\cdots c_M^\dagger) ,
\label{eq:gamma}
\end{equation}
where $A$ is the $M\times M$ matrix representation $A_{lm}=\langle l|\hat A |m\rangle$ 
of operator $\hat A$ on 
a single particle Hilbert space of spinless fermions $\{|l\rangle \}=\{c_l^\dagger|0\rangle \}$, 
with a creation operator, $c^\dagger_l$ ($l=1\cdots M$), applied on a vacuum $|0\rangle$. 
For two operators, $\hat A$ and $\hat B$, 
we find $[\hat \Gamma(\hat A),\hat \Gamma(\hat B)]= \hat \Gamma([\hat A, \hat B])$. 
We know from the Baker-Campbell-Hausdorff formula that for given matrices $A$ and $B$, 
there is a matrix $C$ that fulfills $\e^A\e^B = \e^C\;$\cite{footnote}. 
Then, the corresponding Fock operators are also given as
\begin{equation}
\e^{\hat\Gamma(\hat A)} \e^{\hat\Gamma(\hat B)}=\e^{\hat\Gamma(\hat C)}.  
\label{eq:kilch}
\end{equation}
As we see shortly, this relationship guarantees that one can rewrite Eq.(\ref{eq:trmatvi}) in a 
single exponential form $\e^{\hat H_i}$ as far as the operators $\hat H_{1;i}, \hat H_{2;i}, \hat H_{2;i+1}$ 
are written by the common single-particle basis. 
\par
Any matrix $C$ can be transformed to 
$P^{-1} C P= {\rm diag}(\xi_1,\xi_2,\cdots \xi_M) + D$ with $D$ being an upper triangular matrix, 
where we are familiar with $D=0$ for the symmetric matrix $C$. 
Since $\hat\Gamma(\hat C) = \sum_m \xi_m d_m^\dagger d_m +\sum_{i<j} D_{ij} d_i^\dagger d_j$ 
with $d_i= \sum_l P^{-1}_{il}c_l$, 
the trace of Eq.(\ref{eq:kilch}) is evaluated as 
\begin{eqnarray}
{\rm Tr}\big( \e^{\hat\Gamma(\hat C)} \big)&=& {\rm Tr}\big(\e^{\sum_m \xi_m d_m^\dagger d_m}\big)
\nonumber \\
&=& {\rm det}\big(I+\e^{{\it diag}(\xi_m)}\big) = \prod_{m=1}^M (1+\e^{\xi_m}) \nonumber \\
&=&{\rm det}\big(I+ \e^{A} \e^{B} \big)
\label{eq:kilch2}
\end{eqnarray}
This trace formula holds for more than two products of exponentials of the operators. 
We apply this formula in obtaining $Z$ for Case (i). 

\subsection{Exact solutions of the deformed 2D Ising models}
\label{sec:ssdcases} 
\subsubsection{Case (i): when the Hamiltonian is uniform along the column and non-uniform along the row}
\label{sec_casei}
In this subsection, 
the parameters $K_{1;i,j}$ and $K_{2;i,j}$ depend on $i$ but not on $j$, 
where we omit index-$j$ for simplicity. 
The formula Eqs.(\ref{fr})-(\ref{eq:h1qh2q}) still holds. 
However, since the transfer matrix $V_i$ depends on column-index $i$, and 
since $V_i$ is no longer symmetric, we cannot straightforwardly prepare 
an orthogonal matrix $P$ that diagonalizes all $V_i$'s simultaneously. 
\par
Instead of dividing $Z$ into columns, we first take the whole product over the columns 
for each $q$ to obtain $\hat V_q$, and then combine all $q$ sectors as 
\begin{eqnarray}
&& Z= \prod_{i=1}^L (2\sinh 2K_{1;i})^{L/2} {\rm Tr}\big( \prod_q \hat V_q \big)\nonumber \\
&& \hat V_q=\prod_{i=1}^L \e^{\hat H_{2;iq}} \e^{\hat H_{1;iq}} \equiv \e^{\hat H_q}. 
\end{eqnarray} 
We again find a final single exponential form of the Fock operator $\hat H_q$ 
since $\hat H_{1;iq}$ and $\hat H_{2;iq}$ 
fulfill the condition for $\hat\Gamma(\hat A)$ and $\hat\Gamma(\hat B)$ 
in Eq.(\ref{eq:kilch}). 
By multiplying the $2\times 2$ matrix in Eq.(\ref{eq:h1qh2q}), we obtain 
an explicit form 
\begin{eqnarray}
\e^{\hat H_q}= \bm \eta_q^\dagger (\prod_{i=1}^L \e^{H_{2;iq}} \e^{H_{1;iq}} )  \bm \eta_q. 
\label{eq:hq_casei}
\end{eqnarray}
This matrix is diagonalized to ${\rm diag}(\e^{-2E_q},\e^{2E_q})$
by performing a Bogoliubov transformation $(\eta_q^\dagger,\eta_{-q}) \rightarrow (\tilde \xi_q^\dagger, \tilde \xi_{-q})$ 
similarly to Eq.(\ref{eq:bgtr}), and we find a final form 
\begin{eqnarray}
&& \hat V_q= {\rm exp}\bigg(
\sum_{0\le q <2\pi} -2E_q \,\big(\tilde \xi_q^\dagger \tilde \xi_q  -\frac{1}{2} \big)\bigg). 
\label{eq:hq_casei}
\end{eqnarray}
Here, $E_q \ge 0$ is an order-$L$ quantity, which is the energy carried by 
the Bogoliubov quasi-particle $\tilde \xi_q$. 
In obtaining $E_q$, fermions, $\eta_q$ and $\eta_{-q}$, mix for $0<q<\pi$ 
so that the operator in Eq.(\ref{eq:hq_casei}) is represented by a $2\times 2$ matrix. 
Whereas for $q=0$ and $\pi$, there is no mixing 
and we obtain $E_0=\sum_i (K_{1;i}^* - K_{2;i})$ and $E_\pi= \sum_i(K_{1;i}^*+ K_{2;i})$. 
\par
The partition function is obtained as
\begin{eqnarray}
Z&=& \big(\prod_{i=1}^L (2\sinh 2K_{1;i})^{L/2}\big) \prod_{0\le q<2\pi} \big(\e^{E_q}+ \e^{-E_q} \big)
\rule{3mm}{0mm}
\end{eqnarray}
Among the contributions from Bogoliubov quasi-particles to $Z$, namely the last product term in the 
above equation, a so-called largest eigenvalue is obtained solely from 
a ``vacuum" state of Bogoliubov quasi-particle as 
\begin{eqnarray}
&& \Lambda_{0}= \big(\prod_{i=1}^L (2\sinh 2K_{1;i})^{L/2}\big) \prod_q \e^{E_q}, 
\label{lamq_max}
\end{eqnarray} 
and for $L\rightarrow\infty$ we obtain $Z=\Lambda_0$. 
Exciting a single Bogoliubov quasi-particle with a minimum excitation energy, $2{\rm min}(E_q)$, 
yields the next-largest eigenvalue, so that $\Lambda_{1}/\Lambda_{0}=\e^{-2{\rm min}(E_q)}$. 
The rest of the eigenvalues are determined by successively exciting quasi-particles 
$\prod_{q\in \{q\}_m} \xi_q^\dagger \cket{0}$, where $\{q\}_m$ are a sets of indices of excited particles. 
\par
The energy densities per bond along the column and row are formally given as
\begin{eqnarray}
\langle \sigma_{i,j}\sigma_{i,j+1}\rangle 
&=& - \frac{1}{L} \frac{\partial \ln Z}{\partial K_{2;i}}
= -\frac{1}{L}\sum_{0\le q<2\pi} \frac{d E_l}{dK_{2;i}}
\nonumber \\
\langle \sigma_{i,j}\sigma_{i+1,j}\rangle 
&=& -\frac{1}{L}\frac{\partial \ln Z}{\partial K_{1;i}}
\nonumber\\ 
&=& - \frac{1}{\tanh 2K_{1;i}} 
-\frac{1}{L}\sum_{0\le q<2\pi} \frac{d E_l}{dK_{1;i}}. 
\end{eqnarray}
In numerically evaluating these quantities, the derivatives are much less accurate than those 
we obtain for Case (ii) in the next section. 
%
\subsubsection{Case (ii): when the Hamiltonian is non-uniform along the column and uniform along the row}
\label{sec_caseii}
In this subsection, we consider Case (ii). 
We start from Eqs.(\ref{v1hams})-(\ref{eq:trmatvi}). 
From the discussions given in \S \ref{sec_klich}, 
the transfer matrix operator in Eq.(\ref{eq:trmatvi}) 
is rewritten as $\hat V_i=\e^{\hat H_{2;i}/2}\e^{\hat H_{1;i}}\e^{\hat H_{2;i}/2} =\e^{\hat H_{i}}$, 
where $\hat H_{i}$ takes a quadratic form of a set of one-body operators $\{\bm c_i \}$. 
Since $\hat V_i$'s do not depend on a column-index $i$, 
their representations are separately diagonalized simultaneously for all columns. 
This time, however, the Hamiltonian is non-uniform along the column. 
Then, the $L\times L$ matrix representation of $\hat H_{i}$ 
can no longer be block diagonalized into smallest pieces by the Fourier transformation, 
nor can we apply a simple Pauli matrix representation we used in obtaining Eq.(\ref{eq:viuniform}). 
Instead, the form of $\hat H_{i}$ is obtained through the following processes. 
\par 
We first describe $\hat H_{1;i}$ and $\hat H_{2;i}/2$ in a quadric form of the one-body operators. 
Since the number of fermions does not conserve, we need to prepare a set of $L$-independent 
creation and annihilation operators. 
We reduce the $2L$ operators $\{ c_j^\dagger, c_j \}$ ($j=1,\cdots,L$) by half to avoid redundancy\cite{footnote2}. 
For this purpose, we use the reflection symmetry of $f_{\rm SSD}$ about the center of the system, $K_{2;ij}=K_{2,iL-j+1}$. 
The operators that fulfill ${\cal M}^{-1}a_j{\cal M} =-a_j$ and ${\cal M}^{-1}b_j{\cal M} =b_j$ 
about the parity operator ${\cal M}$ of the mirror reflection are 
\begin{eqnarray}
&& a_j^\dagger = \frac{1}{\sqrt{2}} (c_j^\dagger - c_{L-j+1}^\dagger), \nonumber \\
&& b_j^\dagger = \frac{1}{\sqrt{2}} (c_j^\dagger + c_{L-j+1}^\dagger), \;\; (j=1 \,\sim \,\frac{L}{2}). 
\end{eqnarray}
By using $\bm \Phi^\dagger=(b_1^\dagger ,\cdots,b_{\frac{L}{2}}^\dagger, a_1,\cdots a_{\frac{L}{2}})$, 
the following expressions are obtained; 
\begin{eqnarray}
&& \hat H_{1;i} = \sum_{j=1}^{\frac{L}{2}} -2 K_{1;ij}^* \big( b_j^\dagger b_j - a_j a_j^\dagger \big),
\label{eq:h1i_case2}
\\
&& \frac{\hat H_{2;i}}{2} = \bm \Phi^\dagger
\left( \begin{array}{ll}
\;\;A^+ & \;\;B^- \\
-B^+ &  -A^-
 \end{array}\right)
\bm \Phi \:- {\rm Tr}(A^+) + {\rm Tr}(A^-), 
\nonumber\\
&& A^\pm_{mn}=\frac{K_{2;im}}{2}  \delta_{m+1,n}+ \frac{K_{2;in}}{2} \delta_{m-1,n}
\pm \frac{K_{2;i\frac{L}{2}}}{2} \delta_{m \frac{L}{2}}\delta_{n \frac{L}{2}},
\nonumber\\
&& B^\pm_{mn}=\frac{K_{2;im}}{2}  \delta_{m+1,n}- \frac{K_{2;in}}{2} \delta_{m-1,n}
\pm \frac{K_{2;i\frac{L}{2}}}{2} \delta_{m \frac{L}{2}}\delta_{n \frac{L}{2}}, 
\nonumber\\
\end{eqnarray}
where $A^\pm$ and $B^\pm$ are the $L/4 \times L/4$ matrices. 
\par
Next, we find a matrix $Q$ to transform $\bm \eta$ for $\hat H_{2;i}/2$ as 
\begin{eqnarray}
&& \bm \eta^\dagger = 
(\eta_1^\dagger, \cdots, \eta_{L}^\dagger)
= \bm \Phi^\dagger Q, 
\\
&&\frac{\hat H_{2;i}}{2} = \sum_{l=1}^{L} 
\gamma_l\; \eta_l^\dagger \eta_l. 
\end{eqnarray}
Here, $\gamma_l$ with $l=1\sim \frac{L}{2}$ are nonnegative and 
the other half with $\frac{L}{2}+1\sim L$ are nonpositive. 
Since $\e^{\hat H_{2;i}/2}= \sum_l \e^{\gamma_l} \eta_l^\dagger \eta_l$, 
one can put this back to the original representation as 
\begin{equation}
\e^{\hat H_{2;i}/2}=  \bm \eta^\dagger \e^{{\rm diag}(\gamma_l)} \bm \eta 
=\bm \Phi^\dagger Q \e^{{\rm diag}(\gamma_l)} Q^{-1} \bm \Phi. 
\end{equation}
We thus obtain the Fock operator 
\begin{eqnarray}
&& \e^{\hat H_i}\equiv  \bm \Phi^\dagger T \bm \Phi, \nonumber \\
&& T= Q \e^{{\rm diag}(\gamma_l)} Q^{-1} \e^{H_{1;i}} Q \e^{{\rm diag}(\gamma_l)} Q^{-1} ,
\label{eee}
\end{eqnarray}
where from Eq.(\ref{eq:h1i_case2}), 
$(\e^{H_{1;i}})_{lm}= \delta_{lm} \e^{-2K_{1;ij}^*}$ for $l \le \frac{L}{2}$ 
and $\delta_{lm} \e^{2K_{1;ij}^*}$ for $\frac{L}{2}+1 \le l$. 
\par
As a third step, we diagonalize Eq.(\ref{eee})  as $P^{-1} T P= {\rm diag}(\zeta_l)$, 
by a Bogoliubov transformation 
$\bm \xi^\dagger =(\xi^\dagger_1,\cdots, \xi^\dagger_{\frac{L}{2}},\bar\xi_{1},\cdots, \bar\xi_{\frac{L}{2}})= \bm \Phi P$. 
Here, the distribution of eigenvalues is such that half of $\ln\zeta_l$ are nonpositive
and the other half are nonnegative. 
Therefore, by setting $\zeta_l$ in ascending order 
and by putting it back to the exponential form with $2\epsilon_l=|\ln\zeta_l|$, we find 
\begin{eqnarray}
&& \hat V_{i} = \big(\prod_{j=1}^{L} (2\sinh 2K_{1;ij})^{1/2}\big) \e^{\hat H_{i}},
\nonumber\\
&& \hat H_{i}= 
\sum_{l=1}^{\frac{L}{2}} -2\epsilon_l \xi_l^\dagger \xi_l 
+ 2\epsilon_{m} \bar\xi_m \bar\xi_m^\dagger
= \sum_{l=1}^L -2\epsilon_l \big( \xi_{l}^\dagger \xi_{l}-\frac{1}{2} \big),\rule{5mm}{0mm}
\label{hssdfinal}
\end{eqnarray}
where for $m=l+\frac{L}{2}$ we have $\ln\zeta_m\ge 0$. 
We also applied a particle-hole transformation 
$\bar\xi_m\bar\xi_m^\dagger=1-\xi_m^\dagger \xi_m$. 
From Eq.(\ref{eq:kilch2}), we find 
\begin{eqnarray}
{\rm Tr}\big( \prod_{i=1}^L \e^{\hat H_i} \big) 
&=& \e^{L\sum_l \epsilon_l}{\rm det}\Big( I + \big(\e^{{\rm diag} (-2\epsilon_l)}\big)^L \Big)
\nonumber\\
&=& \prod_{l=1}^L \big(\e^{\epsilon_l L }+ \e^{-\epsilon_l L}\big).
\end{eqnarray}
The partition function is obtained from Eq.(\ref{eq:kilch2}) as
\begin{eqnarray}
Z&=& \prod_{j=1}^L (2\sinh 2K_{1;ij})^{L/2}
 \prod_{l=1}^L \big(\e^{\epsilon_l L }+ \e^{-\epsilon_l L}\big) 
\nonumber \\
&=& \sum_{m=0}^{2^L-1} \Lambda_m
\rule{3mm}{0mm}
\end{eqnarray}
with $\Lambda_0=\prod_{j=1}^L (2\sinh 2K_{1;ij})^{L/2} \prod_{l=1}^L \e^{\epsilon_l L}$, 
being a partition function at $L\rightarrow\infty$. 
\par
The bond energy along the column and the row are given as 
\begin{eqnarray}
\langle \sigma_{i,j}\sigma_{i,j+1}\rangle 
&=& -\sum_{l=1}^L \tanh (L\epsilon_l) \frac{d\epsilon_l}{dK_{2;ij}}
\label{ss:caseii-1}
\\
\langle \sigma_{i,j}\sigma_{i+1,j}\rangle 
&=& - \frac{1}{\tanh 2K_{1;ij}} 
- \sum_{l=1}^L \tanh (L\epsilon_l) \frac{d\epsilon_l}{dK_{1;ij}}.
\label{ss:caseii-2}
\end{eqnarray}

%
\begin{figure*}[tbp]
  \begin{center}
   \includegraphics[width=15cm]{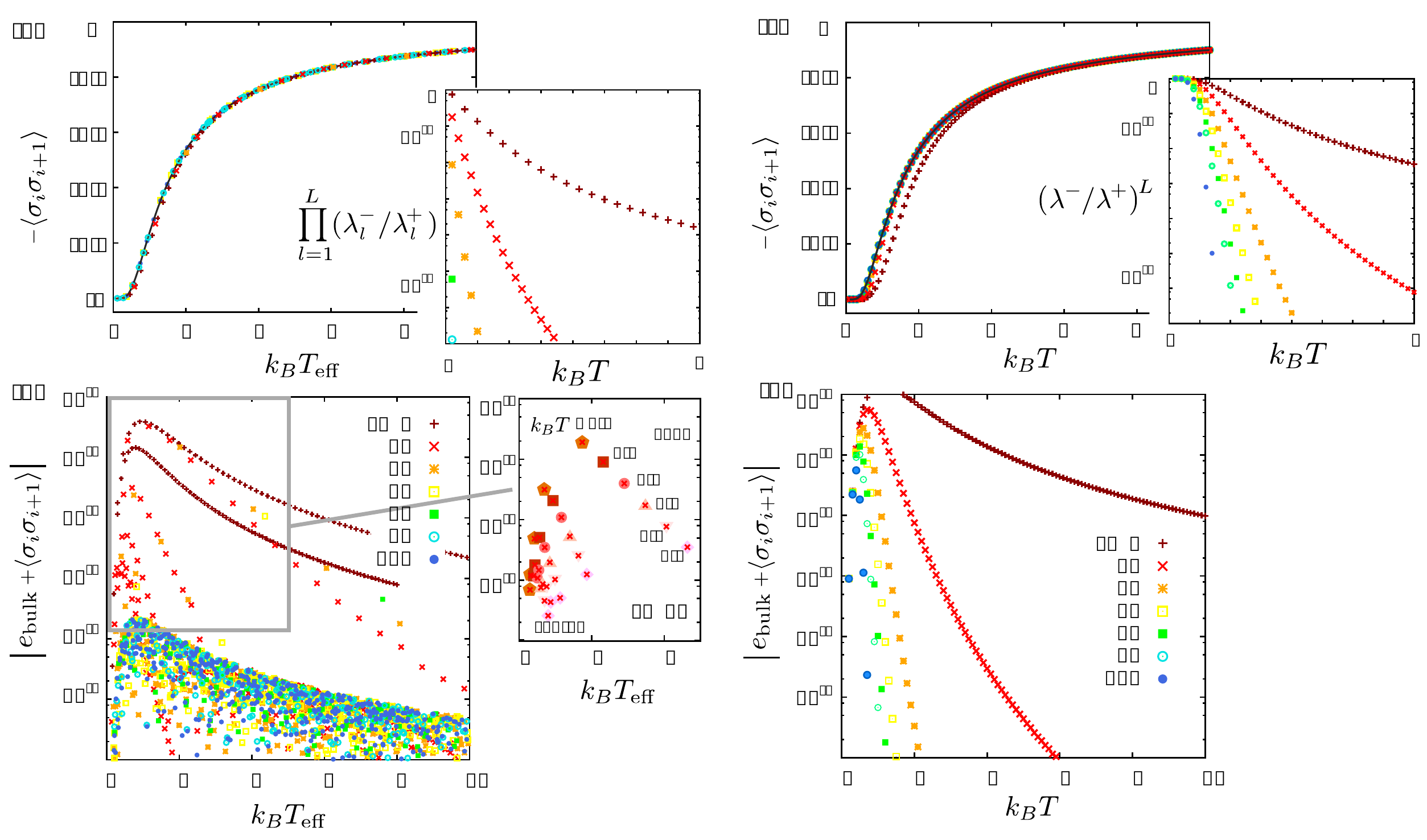}
  \caption{Panels (a,b) are the results of the SSD Ising model in one dimension, 
   and panels (c,d) are those of the uniform Ising model to be compared with (a,b). 
   (a) Location (index-$i$) dependent bond energy, $-\langle \sigma_i\sigma_{i+1}\rangle$, 
    given in Eq.(\ref{bond1d}) plotted against the effective temperature $k_BT_{\rm eff}$ 
    for $L=4 \sim 100$. 
    Solid line is the exact energy of a uniform Ising model $e_{\rm bulk}$ at $L=\infty$ 
    as a function of temperature. 
   (b) Deviations of bond energy in (a) from $e_{\rm bulk}$. 
    Calculations are done for several choices of $k_BT=0.1n$ ($n$:integer), 
    where each single choice of $k_BT$ generates $L/2$ data points. 
    Inset shows how the results vary with $n$ for $L=10$ magnified from the main panel; 
    the better accuracy is observed for the bonds closer to the system center. 
  (c) Exact bond energy for $L=4\sim 100$ in the uniform 1D Ising model, 
  and (d) the finite-size correction against $e_{\rm bulk}$. 
  Insets of (a) and (c) show $(\lambda^-/\lambda^+)^L$ as a function of temperature. 
}
    \label{f3}
  \end{center}    
\end{figure*}
%
\section{Numerical examination}
\label{sec_numerical}
In this section, we numerically demonstrate how SSD works on the Ising model by using 
the formula in the previous section. 
We also compare these results with those of classical Monte Carlo simulations. 
\subsection{1D systems} 
In \S\ref{1dtrmat}, we found that the maximum eigenvalue of the transfer matrix 
on the bond at position $\bm r_i$ connecting site $i$ and $(i+1)$ 
serves as a local partition function on that bond, 
\begin{equation}
\lambda_i^+=\e^{-J/k_BT_{\rm eff}(r_{i}) }+ \e^{J/k_BT_{\rm eff}(r_{i})}, 
\label{eq:z1d}
\end{equation}
at its effective temperature, $k_BT_{\rm eff}(r_{i})=k_BT/f_{\rm SSD}(r_{i})$. 
In the thermodynamic limit, the second largest eigenvalue $\lambda_i^-$ is neglected 
and the total partition function becomes a product of $\lambda_i^+$. 
This means that the system is an ensemble of $(L-1)$-different noninteracting bond degrees of freedom, 
and unlike a uniform system, each is exposed to its own temperature $k_BT_{\rm eff}$ that depends on its location. 
The form of Eq.(\ref{eq:z1d}) indicates that one can obtain a set of equilibrium states with different temperatures $k_BT_{\rm eff}$ ranging from $k_BT$ to $\infty$, simultaneously, in a single system. 
\par
To examine how accurate the above mentioned description would be at finite $L$, 
we numerically evaluate the bond energy of the SSD Hamiltonian in Eq.(\ref{bond1d}) 
as a function of $k_BT_{\rm eff}$ for several choices of $L=4,\cdots,100$, which is shown in Fig.~\ref{f3}(a). 
The bond energy even at $L=4$ shows relatively good agreement with the exact bond energy $e_{\rm bulk}=-\tanh (\beta J)$ 
of the $L=\infty$ uniform Ising model. 
The inset of Fig.~\ref{f3}(a) shows $\prod_i(\lambda_i^-/\lambda_i^+)$, 
which is a rapidly decreasing function of both $k_BT$ and $L$. 
When this quantity is sufficiently small, $Z=\prod_i \lambda_i^+$ holds, and $\lambda_i^+$ 
given in Eq.(\ref{eq:z1d}) serves as a local partition function for the corresponding local effective temperature, 
which is fulfilled for most of the temperature range $k_BT \gtrsim {\cal O}(0.1J)$. 
Therefore, one can realize a canonical ensemble of systems with a variety of temperatures in a single system by properly setting $k_BT$. 
\par
Figure~\ref{f3}(b) shows the deviation of bond energy against $e_{\rm bulk}$, which we call an SSD error. 
For $L\gtrsim 40$, it is less than 10$^{-5}$. 
Here, a single partition function for a fixed $k_BT$ generates $L/2$-independent data points with different $k_BT_{\rm eff}$. 
Therefore, various $L/2$-sets of data are obtained by varying $k_BT=0.1n$ with $n=1,2,\cdots$. 
The inset shows the SSD error of these series for $L=10$; 
the data obtained near the system center has better accuracy than those near the edges. 
We also found that the accuracy is improved for higher $k_BT$. 
We may explain this tendency by a slope of the effective temperature. 
The slope is gentle at the center and becomes steeper in approaching the edge of the system as shown in Fig.~\ref{f1}(a). 
The location that gives a certain fixed value of the effective temperature becomes closer to the system center 
if the system temperature $k_BT$ is larger. 
The local thermal equilibrium is better attained if an additional energy flow caused by the slope of the temperature can be neglected. 
\par
The results presented above are in good agreement with the tendency observed in the quantum many-body systems under the SSD. 
In the quantum cases\cite{hotta12}, the physical quantities at $T=0$ measured at the system center reproduce the values in the infinitely large system even when the system is as small as $L\lesssim 20$. 
At finite temperature, it works quite well even at $L=4$ in the 1D system \cite{chisa18}. 
In the same context, the measurements of $-\langle \sigma_i\sigma_j\rangle$ in the present classical system is accurately performed near the system center by varying $k_BT$ even at small $L$. 
Generally, it is easy to increase $L$ by one order of magnitude in classical systems, 
and the measurements over a wide range of system become accurate enough (see Fig.\ref{f3}(b)). 
This fact will be of great advantage in utilizing the SSD for Monte Carlo simulations. 
\par
In Figs.~\ref{f3}(c) and \ref{f3}(d), we show the results of a {\it uniform} Ising model at finite $L$ 
to compare with Figs.~\ref{f3}(a) and \ref{f3}(b), respectively. 
When $\lambda^-/\lambda^+$ shown in the inset becomes small enough, the bond energy approaches $e_{\rm bulk}$. 
In contrast to the case of SSD, the bond energy at $L=4$ disagrees with $e_{\rm bulk}$ by about 10$^{-2}$. 
An advantage of the SSD system over the uniform system is particularly significant at around $k_BT_{\rm eff}\sim 1$. 
The finite-size correction of the uniform system 
remains of order $10^{-2}$ even when increasing $L$ up to 100, 
where we find $\lambda^-/\lambda^+ \sim 1$ accordingly.

\begin{figure*}[tbp]
  \begin{center}
    \includegraphics[width=18cm]{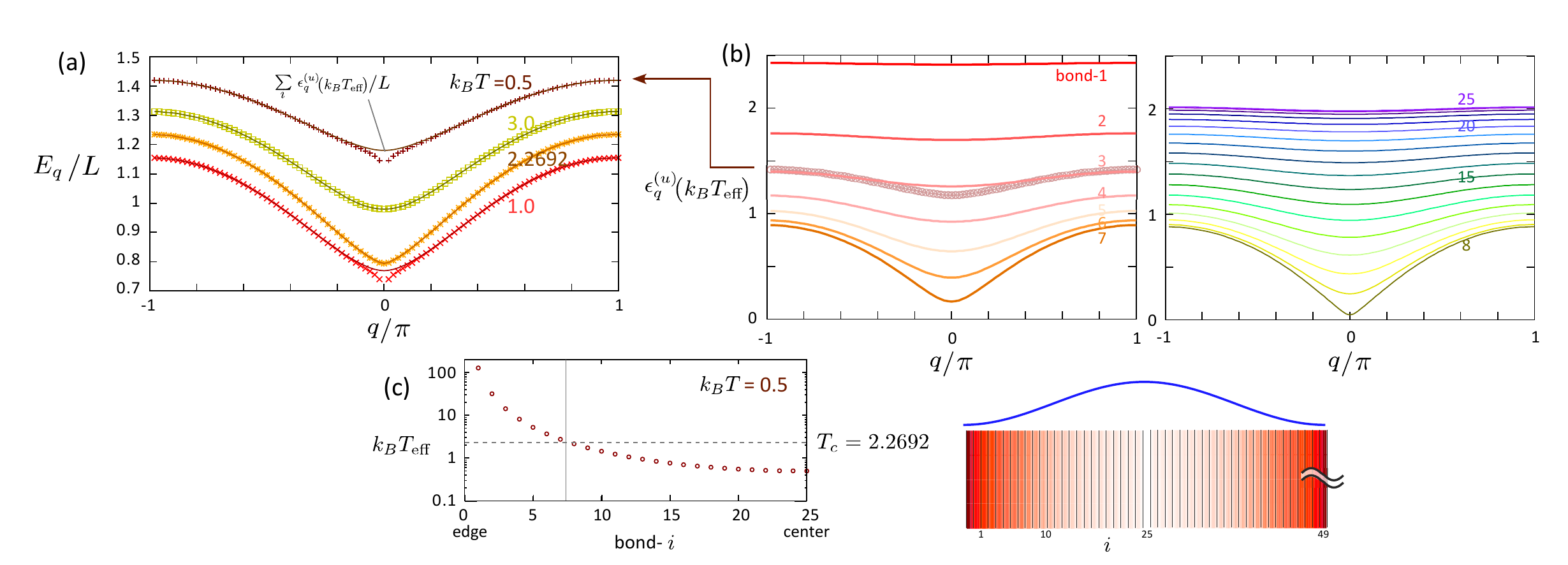}
	  \caption{ Dispersions of fermions of a transfer matrix along the column ($j$-direction) with $L=50$ in Case (i). 
	  (a) $E_q/L$ in Eq.(\ref{eq:hq_casei}) for Case (i) with $L=50$ (symbols) 
	     at $k_BT=0.5,1,2.2692(T_{c})$ and 3. 
     (b) $\epsilon_q^{(u)}$ obtained using Eq.(\ref{eq:epsilon}) using the temperature 
     $k_BT_{\rm eff}$ that depends on the location of bond-$i$, $i=1 \sim 25$.
     (c) The $i$-dependent $k_BT_{\rm eff}$ used to evaluate the dispersions in (b). 
     Solid lines in (a) is the summation of $\epsilon_q^{(u)}$ throughout the system, $i=1\sim 50$, 
     to be compared with $E_q$ for the same system temperature $k_BT$. 
  The right panel is the density plot of $k_BT_{\rm eff}$, 
  together with the profile of $f_{\rm SSD}$ along the rows. 
   }
    \label{f4}
  \end{center}    
\end{figure*}
\begin{figure*}[tbp]
  \begin{center}
   \includegraphics[width=18cm]{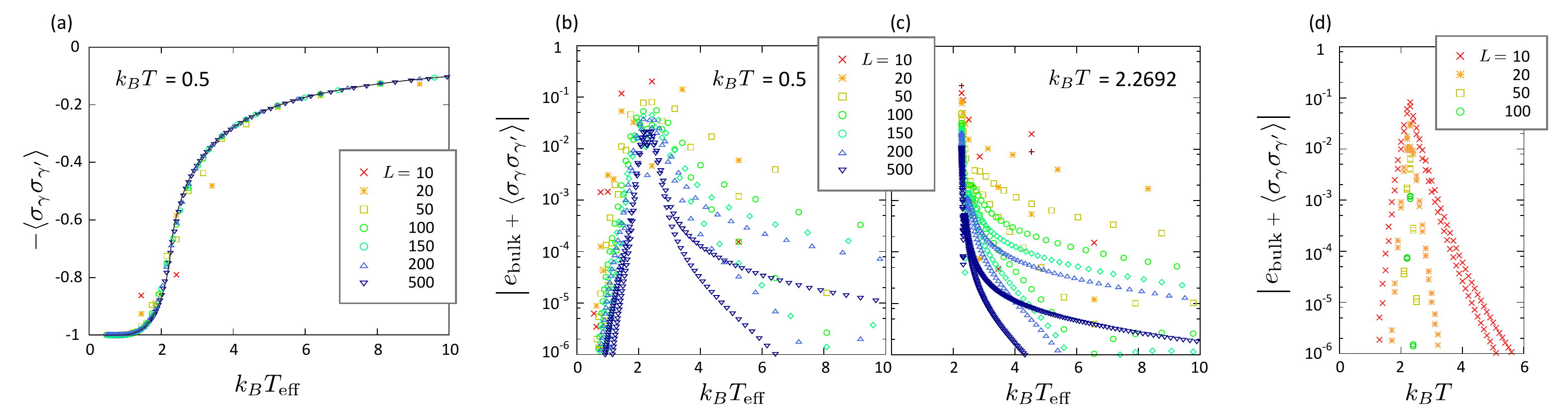}
  \caption{(a) Bond energy $-\langle \sigma_{\gamma} \sigma_{\gamma'}\rangle$ 
  evaluated for the 2D SSD Ising model plotted against $k_BT_{\rm eff}$ 
  using Eqs.(\ref{ss:caseii-1}) and (\ref{ss:caseii-2}) for Case (ii) 
  with $L=10\sim 500$ and $k_BT=0.5$. 
  Solid line is the exact bulk energy, $\epsilon_{\rm bulk}$. 
  (b) SSD error $|e_{\rm bulk}+\langle \sigma_{\gamma} \sigma_{\gamma'}\rangle|$ for the data in panel (a) with $k_BT=0.5$. 
  (c) SSD error for a set of bond energy evaluated for $k_BT=2.2692=k_BT_{c}$. 
  (d) Finite-size corrections of the bond energy evaluated for 
  the uniform 2D system using Eqs.(\ref{ss:caseii-1}) and (\ref{ss:caseii-2}) as a function of $k_BT$. 
	  }
    \label{f5}
  \end{center}
\end{figure*}
%
\begin{figure}[tbp]
  \begin{center}
   \includegraphics[width=9cm]{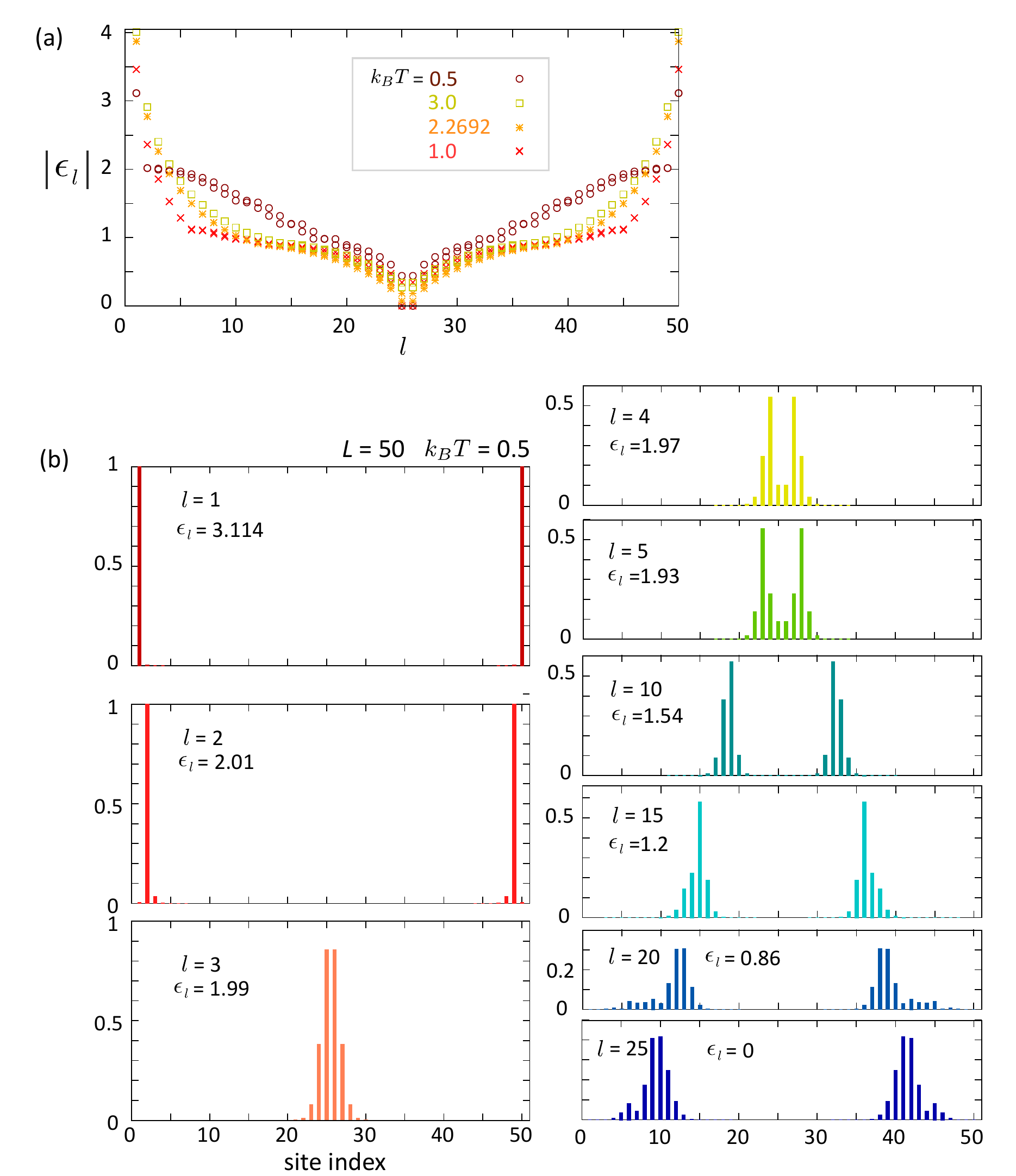}
    \caption{ (a) Dispersions $\epsilon_l$ of fermions of a transfer matrix along the 
  column ($j$-direction) with $L=50$ in Eq.(\ref{hssdfinal}) for Case (ii) 
  with $k_BT=0.5,1,$ 2.2692$(T_{c})$ and 3. 
   (b) The spatial amplitude of the one-body eigen state of index-$l$ calculated at $L=50$ and $k_BT=0.5$ 
   in panel(a). The bottom panel $l=25$ state is the zero energy state.
}
    \label{f6}
  \end{center}    
\end{figure}
%
\subsection{2D systems} 
\label{ssdresults}
We apply the SSD along the row in Case (i) and along the column in Case (ii) (see Fig.\ref{f1}(c)). 
The fermionic dispersions along the column, $E_q/L$ in Case (i) 
and $\epsilon_l$ in Case (ii), are presented. 
For the SSD Hamiltonian, these two cases give different energy dispersions, 
since for Case (i) $q$ is a good quantum number along the column but for Case (ii) it is not. 
We checked that when $J_{ij}$ is uniform, Cases (i) and (ii), whose Hamiltonian is the same but the formulation differ, give the same results. 

\subsubsection{Fermionic dispersions}
Let us first compare the energy dispersions of fermions as functions of $q$ in the SSD and the uniform systems. 
The purpose here is to examine whether 
the partition function $Z\rightarrow \Lambda_0$ at $L\rightarrow\infty$ 
for the SSD in Case (i) given in Eq.(\ref{lamq_max}) is 
equivalent to the product of local partition function of each column with different $i$-dependent $k_BT_{\rm eff}(\bm r_i)$. 
The latter is formally obtained by replacing the uniform temperature $k_BT$ of $\Lambda_0$ with the $i$-dependent 
$k_BT_{\rm eff}(\bm r_i)$ in Eq.(\ref{eq:z_uniforminfty}) in \S.\ref{sec:uniform}. 
This replacement is equivalent to having the relation 
\begin{equation}
\sum_{i=1}^L \epsilon_q^{(u)} = E_q, 
\label{eq:replace}
\end{equation}
where $\epsilon_q^{(u)}$ of the uniform system 
calculated for each $k_BT_{\rm eff}(\bm r_i)$ is summed over different columns on the l.h.s. 
and $E_q$ on the r.h.s. is obtained in Eq.(\ref{lamq_max}). 
If this equation exactly holds, 
the classical SSD system in 2D is an extended canonical 
ensemble of a local subsystems each in an equilibrium of different temperature. 
Since $E_q$ cannot be obtained analytically, we show numerically that this relationship holds 
almost exactly except for the small deviation at around $q\sim 0$. 
\par
Figure~\ref{f4}(a) shows $E_q/L$ in Eq.(\ref{eq:hq_casei}) for Case (i) with $L=50$ 
for several choices of $k_BT$. 
Since $E_q$ is the order-$L$ quantity obtained after multiplying the transfer matrices of all columns, 
we expect that it can be approximately divided into contributions from different columns 
if these columns can be regarded as independent subsystems, 
which is the implication of Eq.(\ref{eq:replace}). 
In Fig.\ref{f4}(b) we plot together a set of $L/2$ independent 
dispersions of fermions representing a single transfer matrix 
of a uniform system $\epsilon_q^{(u)}$, each obtained for column(bond-$i$) dependent $k_BT_{\rm eff}$ 
using Eq.(\ref{eq:epsilon}). 
A set of effective temperatures $k_BT_{\rm eff}$ for $k_BT=0.5$ adopted in this calculation is shown in Fig.\ref{f4}(c). 
At $i=1\sim 7$ where $k_BT_{\rm eff} \le k_BT_{c}$ in the left panel, 
the dispersion is a descending function of $i$, and then for $i\ge 8$ in the right panel it ascends with $i$. 
By averaging all the dispersions over $i=1\sim L$, the data points in panel (b) 
(brown symbols in the l.h.s. panel) are obtained, 
which is the same data as the one plotted in solid line in panel (a) marked by arrows. 
We found that except for the very vicinity $q/\pi \sim 0$, 
the average $\sum_i \epsilon_q^{(u)}/L$ and $E_q/L$ are in almost perfect agreement. 
The same comparison holds for other $k_BT$. 
\par
In this way, the contributions to $E_q$ from each column are well approximated by $\epsilon_q^{(u)}$ 
under locally-defined effective temperature $k_BT_{\rm eff}$. 
The result indicates that a picture we proved in 1D also holds in 2D, 
namely, the system can be regarded as an assemblage of 
small subsystems having a different canonical temperature $k_BT_{\rm eff}$. 

\subsubsection{Bond energy}
Since we found 
that the 2D SSD system can simultaneously host $L/2$ different subsystems with different effective temperature, 
we can use this fact to evaluate the local physical quantities in each subsystem. 
Figure~\ref{f5}(a) shows 
bond energy $-\langle \sigma_{\gamma} \sigma_{\gamma'}\rangle$ 
as a function of $k_BT_{\rm eff}$ obtained for Case (ii) using Eqs.(\ref{ss:caseii-1}) and (\ref{ss:caseii-2}), 
where we set $k_BT=0.5$ and plot the results for $L=10\sim 500$. 
Solid line $e_{\rm bulk}$ is an exact solution of $L=\infty$-uniform Ising model. 
The data at $L=10$ still deviate from $e_{\rm bulk}$ but 
when $L\gtrsim 50$ they almost perfectly overlap with $e_{\rm bulk}$. 
The SSD error $|e_{\rm bulk}+\langle \sigma_{\gamma} \sigma_{\gamma'}\rangle|$ 
for the same data set is shown in Fig.~\ref{f5}(b). 
There are two series of data points following different curves for the same value of $L$.
One is the bond energy evaluated along the columns, and the other one is along the rows. 
The error is suppressed to less than $10^{-3}$ in a wide range of $k_BT_{\rm eff}$ when $L\gtrsim 100$. 
Setting $k_BT$ to $k_BT_{c}$ further suppresses the SSD error as shown in Fig.~\ref{f5}(c).  
This is because the spatial slope of the effective temperature becomes gentle at the critical temperature $k_BT_c$ where the finite-size effect is very strong.   
The local subsystem can more easily attain the thermodynamic equilibrium at the target effective temperature when the differences of $k_BT_{\rm eff}$ with its neighbors are smaller. 
For comparison, we also calculate the bond energy of a uniform system 
and plot a finite size error in Fig.~\ref{f5}(d). 
They are calculated using Eqs.(\ref{eq:siguniform1}) and (\ref{eq:siguniform2}). 
Near $T_c\sim 2.2692J$, correlation length diverges and the finite size effect becomes large. 
This fact is consistent with the peak of SSD error near $k_BT_{\rm eff}$.

\subsubsection{Eigen states of fermions}
We now examine the spatial distribution of wave functions of fermions when Case (ii) SSD is applied. 
Figure~\ref{f6}(a) shows the energy levels $\epsilon_l$ 
of fermions obtained using Eq.(\ref{hssdfinal}). 
We plot the data for several choices of $k_BT$. 
Since the column direction is no longer uniform, the label $l$ is not a wave number but an index 
in a descending order of $\epsilon_l$ 
for $l=1\sim 25$. The latter half, $l=26\sim 50$, 
takes the same value with the former half due to the reflection symmetry. 
\par
Following Refs.[\onlinecite{maruyama11}] and [\onlinecite{hotta12}], 
we first explain how the SSD term reorganizes the eigen states of fermions. 
By introducing $f_{\rm SSD}(j)=1-g(j)$, which fulfills $g(j)=g(L+1-j)$,  
\begin{eqnarray}
&& g(j)=\cos\bigg( \frac{2\pi}{L}\big(j-\frac{1}{2}\big)\bigg)=g_1\e^{\im \delta j}+ g_{-1}\e^{-\im\delta j} 
\nonumber\\
&& g_1=g_{-1}^*=e^{\im\delta/2} /2 ,
\end{eqnarray}
where $\delta=2\pi/L$, 
one could separate the exponent of Eq.(\ref{v2hams}) into two parts as, 
\begin{eqnarray}
\hat H_{2;i} &=&  \sum_q \hat H_{2;iq}  - \sum_\pm \hat H_{d\pm} 
\label{h2ssd}
\\
\hat H_{d\pm} &=& g_{\pm 1}\Big( \sum_{q} 2\cos\big(q \mp \frac{\delta}{2}\big) \eta_q^\dagger \eta_{q\mp\delta}, 
\nonumber\\
&& + \sum_{0\le q\le\pi} 2\sin \big(q\mp \frac{\delta}{2}\big) \big(\eta_q^\dagger \eta_{-q\pm \delta}^\dagger 
 + \eta_{-q} \eta_{q\mp \delta} \big) \Big).
\label{qmix}
\end{eqnarray}
The first term of Eq.(\ref{h2ssd}) is a $q$-component of Eq.(\ref{heta_q2}). 
In the uniform system, the eigen state is characterized by a wave number $q$. 
By introducing the SSD, one-body states of different values of $q$ mix as in Eq.(\ref{qmix}). 
The particular form of $f_{\rm SSD}$ allows this mixing only between neighboring $q$'s 
which are discretized in a unit of $\delta=2\pi/L$. 
The amplitude of mixing also depends on $q$. 
It takes the largest values at $q=0$ and $\pi$  for the first term, and at $q=\pm \pi/2$ for the second term.
These three are the top or bottom and the middle, 
of the energy band in Fig.~\ref{f6}(a), respectively. 
Such moderate mixing generates a wave packet as an eigenstate of $\hat H_{i;2}$. 
Strictly speaking, the final eigenstates are those of Eq.(\ref{eee}) 
and not of $\hat H_{2;i}$, but once we already have a wave-packet state localized in real space, 
it does not change much by the operation of $e^{H_{1;i}}$, as it has only a diagonal form in the real-space representation. 

In Fig.~\ref{f6}(b), a weight of the one-body state ($\bar \xi_l^\dagger\cket{0}$) at a site-$j$, $|(Q_i)_{jl}|$, 
for the dispersion of the fermions at $k_BT=0.5$ in panel (a) is shown for several energy levels $l$. 
Those labeled by $l=1,2$ are the ones providing the largest two $\epsilon_l$, 
and are almost completely localized at the edge sites. 
This is because the coefficients in Eq.(\ref{qmix}) is 
the largest and it efficiently mixes the states over the wide range of wavenumbers, 
so that the wave packets become a delta function. 
In all energy levels in Fig.~\ref{f6}(b), 
the wave packet typically spans over at most
three to four lattice spacings, and carries a ``bulk" energy $\epsilon_l$, 
which gives a rough characteristic energy scale of $k_BT_{\rm eff}$.
In the uniform and finite systems, the size effect in quantum state occurs because they are plain waves 
with discretized wave number. 
But for SSD, the wave-packet state is localized and does not feel the size of the system. 
Their local physical quantities behave nearly scale free\cite{hotta13}. 
This wave-packet-type distribution of fermions in real space supports the picture that 
the system is an assemblage of local subsystems at different effective temperatures. 
%
\begin{figure*}[tbp]
  \begin{center}
   \includegraphics[width=18cm]{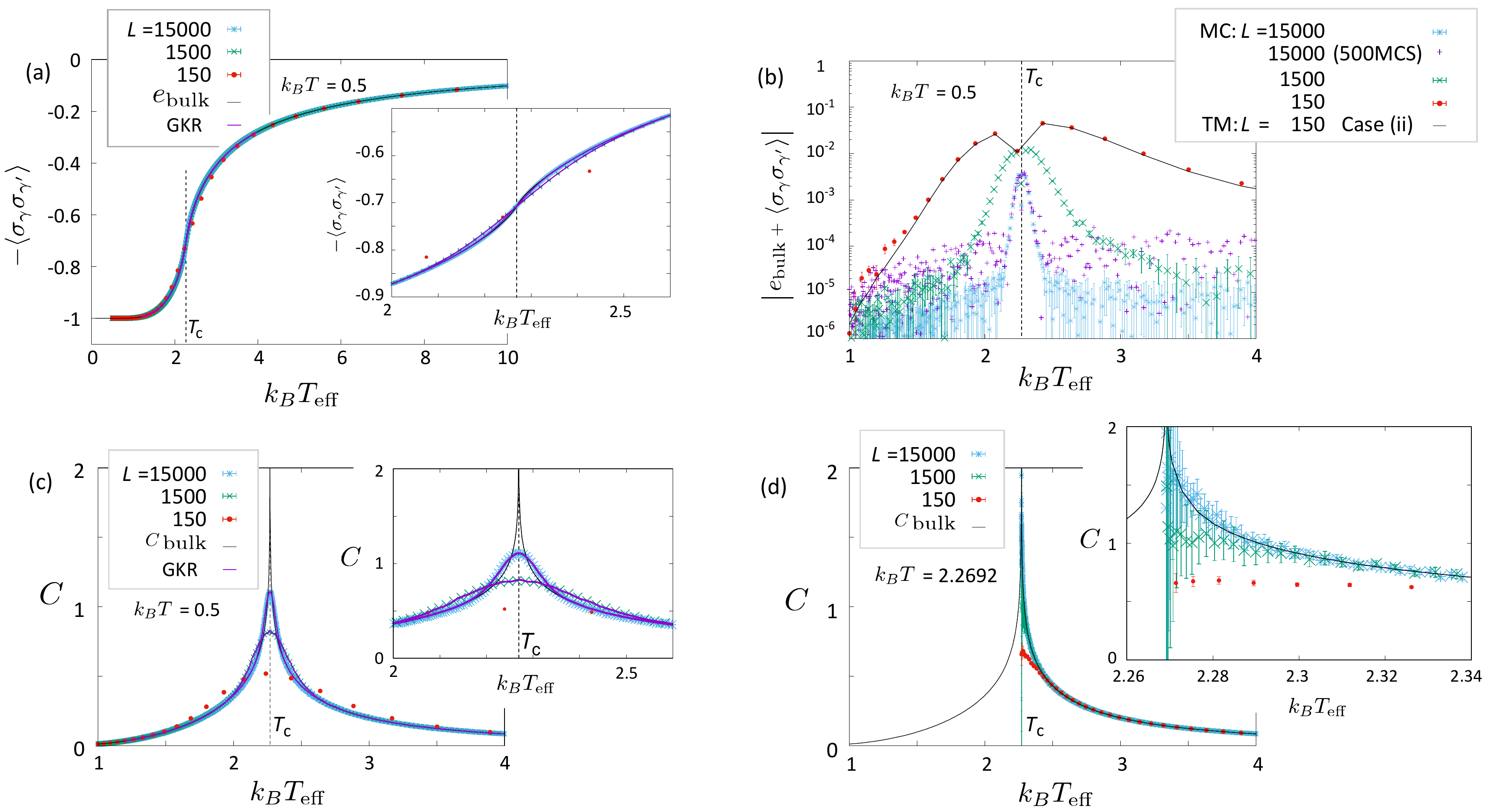}
	  \caption{Results obtained by the classical Monte Carlo simulations 
	  in the 2D Ising model with the SSD along one direction. 
  The simulation temperature is denoted as $k_BT$, 
  which corresponds to the lowest effective temperature. 
  After discarding the initial 5000 steps, we measured the bond energy for 50000 steps unless otherwise denoted.
  Insets show enlarged views near the critical temperature. 
 (a) The bond energy plotted against the effective temperature of each column.
  We compare our results of different lattice sizes with the exact bulk energy, $e_{\rm bulk}$,
  and a result of the Gaussian kernel regression(GKR).
  (b) The SSD error. 
  The MC results of $L=150$ is consistent with the transfer-matrix(TM) results in solid line 
  for Case (ii) SSD of the same size.
  We also plotted a result of $L=15000$ measured for 500 MCS to check
  the MCS dependences.
  (c)-(d) Results of the specific heat compared with the exact bulk value, $c_{\rm bulk}$,
  and the GKR result.
 }
    \label{f7}
  \end{center}    
\end{figure*}

\subsection{Monte Carlo simulation}
\label{sec_mc}
We solved the classical Ising model in {\it a finite system size} {\it exactly} 
using the fermionic representation. 
However, even in the classical systems, the cases with exact solutions are limited. 
A Monte Carlo (MC) simulation usually serves as a good approximate solver.
Applying the SSD to an MC simulation raises a question, 
{\it whether an SSD system converges to a proper equilibrium state even though
the effective temperature of each spin depends on the location? }
We thus carry out the standard single-spin-flip MC simulations to the 2D Ising model deformed in one direction. 
The effective temperature differs for each column as shown in Fig.\ref{f1}(c).
We collect bond energy for each column separately and plot it against the effective temperature. 
We discarded the first 5000 MC steps(MCS) and measured the bond energy for 50000 MCS after that.
We also performed ten independent MC runs and took an average of the data.
The initial spin configuration is the ferromagnetic state with $\sigma_{\gamma}=1$.

Figure~\ref{f7} shows the MC data obtained at different system sizes. 
We checked that the MC result of $L=150$ is consistent with that of the transfer-matrix method shown in Fig.~\ref{f5}. 
As shown in Fig.~\ref{f7}(a), the temperature dependence of the bond energy 
agrees well with the exact results for the whole temperature region.
We find a small discrepancy only in the vicinity of the critical temperature, 
which decreases as the system size increases.
The trend is clearly observed in a plot of the SSD error in Fig.~\ref{f7}(b).
Both a peak value at $T_{\rm eff}=T_{c}$ and the width were found to scale with $1/\sqrt{L}$. 
We also plot in this figure a result of $L=15000$ measured only for 500 MCS after discarding 5000 steps. 
The peak shape at $T_{c}$ is the same as the original measurement with 
50000 MCS but the base of the peak is shifted upward roughly by one order of magnitude ($\sim 10$ times), 
which is the ratio of a square root of two MCS. 
Therefore, the SSD error in this off-critical region is controlled by the standard MC statistical error, 
$1/\sqrt{\rm MCS}\times 1/\sqrt{L}$.
It suggests that the MC approximation would become exact in the limit of an infinite number of steps in this off-critical region of the SSD system. 
On the other hand, the SSD error in the critical region near $T=T_{c}$ is due to the systematic one that solely depends on $L$.
\par
We confirm the validity of the present SSD simulation
by examining how precisely we can reproduce the specific heat from our data.
The specific heat is usually evaluated as a fluctuation of energy based on the two-point correlation of the bond energy over the whole system. 
However, the correlations between different bonds no longer make sense when the SSD is applied.
Instead, we evaluate it from the derivatives of $-\langle \sigma_\gamma \sigma_{\gamma'} \rangle$ 
against $k_BT_{\rm eff}$. 
It is easily performed by a difference between the neighboring effective-temperature data. 
As shown in Fig.~\ref{f7}(c), the specific heat $C$ is also consistent with 
the exact result for the off-critical temperature region.  
The SSD error of the specific heat is consistent with that of the bond energy shown in Fig.~\ref{f7}(b).
\par
Remind that a massive number of temperature data are generated in the SSD simulation 
only by a small numerical effort. 
Taking full advantage of this we apply the Gaussian kernel regression (GKR)
coupled with the Bayesian inference\cite{harada11}. 
The GKR is a machine-learning-based statistical data analysis and with this 
we can estimate the critical temperature, 
and can also obtain bond energy as a continuous function of temperature
without assuming any analytical function form. 
Since a larger number of data sets gives better performance of GKR, 
our SSD MC provides a suitable playground for it. 
In our case, the specific heat is obtained continuously without taking the numerical derivatives.
We only need to take an analytic derivative of the Gaussian distribution function used there.
The regression was already proved to reproduce the critical temperature of the classical 2D Ising model within the accuracy of 10$^{-6}$, and the state-of-art temperature-dependent critical exponent that converges to $\beta=1/8$ at $T=T_{c}$\cite{tota16}. 

We randomly choose 500 data in a range of $1.2\le T \le 4.0$ and apply the GKR
by setting the regression variables $(x_i, y_i)$ as 
$x_i=-\langle \sigma_\gamma \sigma_{\gamma'} \rangle -E_{c}$ and
$y_i=k_BT_{\rm eff} - k_BT_{c}$, where 
$T_{c}$ and
$E_{c}$ (the bond energy at the critical temperature) are 
parameters to be estimated by the Bayesian inference.
Here, we exchange $x_i$ and $y_i$ from the conventional definition because
the bond energy exhibits a singular behavior at $T=T_{c}$.
It is much easier for the regression to model a function with a gentle slope 
than to model a function with a steep slope.
We also know that
the specific heat is symmetric and the bond energy is antisymmetric by a 
mirror reflection of the temperature at the critical temperature in the critical region.
To take this prior information into account,
we introduce a set of mirror data\cite{tota20} 
with respect to the critical point as, $x_i'=-x_i$ and $y_i'=-y_i$.
We mix the data below and above the critical temperature only in the critical temperature region,
$|y_i|<\Delta T$, where $\Delta T$, which is the width of the critical region, 
is another parameter to be estimated by the Bayesian inference. 
The GKR results of $L=1500$ and $L=15000$ are plotted with lines in
Figs.~\ref{f7}(a) and \ref{f7}(c).
The estimated critical temperature and the bond energy values are 
$(T_{c}, E_{c})=(2.2702(3), -0.7072(2))$ for $L=1500$ and 
$(T_{c}, E_{c})=(2.2699(2), -0.7065(2))$ for $L=15000$, whereas
the exact bulk values are
$(T_{c}, E_{c})=(2.2692\cdots, -0.70710\cdots)$.
The width of the critical region was estimated as 
$\Delta T=0.150(7)$ for $L=1500$ and
$\Delta T=0.11(1)$ for $L=15000$.
The critical temperature and the bond energy deviate from the exact results
only by an order of $10^{-4}$. 
Our data in the vicinity of the critical temperature include the SSD error by an order of $10^{-3}\sim -10^{-2}$ as shown in Fig.~\ref{f7}(b).
The difference between the neighboring effective temperatures at $T_{c}$
is more than $10^{-3}$ even in the system of $L=15000$.
The Bayesian inference realizes an accuracy almost ten times
better than these SSD errors.
\par
Since the SSD approximation is generally good at the center of the system,
we can reduce the SSD error in the critical region by setting the 
simulation temperature to the critical temperature 
(in the same context as shown in Figs.~\ref{f5}(b) and \ref{f5}(c)).
Figure~\ref{f7}(d) shows the result of the specific heat.
The error bars near $T_{c}$ are much larger than the result of $k_BT=0.5$. 
The present number of MCS may not be sufficient 
because a real simulation temperature is a critical temperature 
and the critical slowing down may occur. 
In this model, the specific heat diverges at $T_{c}$. 
The exact solution shows that the specific heat reaches $C\sim 2$ 
when the temperature approaches $|T-T_{c}|=4\times 10^{-4}$. 
This exact value is reproduced by the Monte Carlo simulation data for $L=15000$ within the error bar. 
%
\subsection{Other deformation functions}
\label{sec_f}
\begin{figure}[tbp]
  \begin{center}
   \includegraphics[width=9cm]{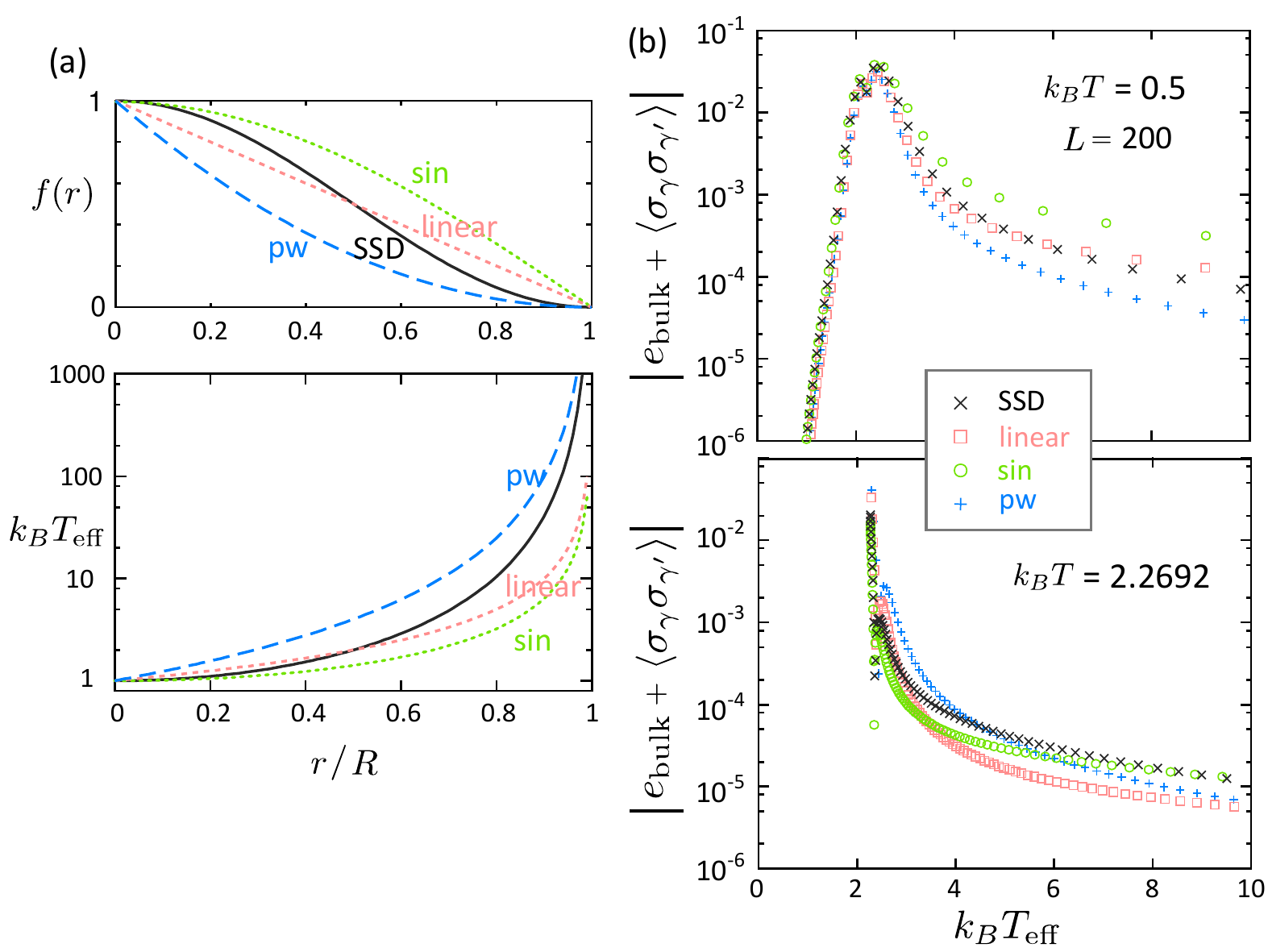}
    \caption{ (a) Functional form of Eq.(\ref{eq:ffunc}) and SSD, and the corresponding 
     effective temperature $k_BT_{\rm eff}$ as functions of $r/R$. 
   (b) The deformation error $|e_{\rm bulk} + \langle \sigma_\gamma \sigma_{\gamma'}\rangle|$ 
       calculated using Case (ii) by replacing the SSD functions with other three functions 
       given in Eq.(\ref{eq:ffunc}). 
       The bond energy for different $k_BT_{\rm eff}$ are obtained at different locations of the system of $L=200$ 
       along the rows. 
}
    \label{f8}
  \end{center}    
\end{figure}

So far we have studied the effect of SSD on classical Ising models. 
However, unlike for the quantum models, our results may suggest that the SSD is not the special deformation for classical models. 
To clarify this point, we performed the same calculation as Case (ii) for other envelope functions. 
We adopt three functions, 
\begin{eqnarray}
&&f_{\rm sin}(r)= \sin\Big(\frac{\pi}{2}(1-\frac{r}{R})\Big) , \\
&&f_{\rm linear}(r)= 1-\frac{r}{R}, \\
&&f_{\rm pw}(r)= \Big(1-\frac{r}{R}\Big)^2, 
\label{eq:ffunc}
\end{eqnarray}
whose spatial dependence and the corresponding $k_BT_{\rm eff}$ as functions of $r/R$ are shown in Fig.~\ref{f8}(a). 
One finds that the slope of $f_{\rm sin}(r)$ and $f_{\rm pw}(r)$ are the decreasing 
and increasing functions of $r/R$ while the slope of $f_{\rm linear}(r)$ is a constant. 
Whereas, the slope of SSD is first a decreasing function and then becomes an increasing function. 
Figure~\ref{f8}(b) shows two panels with different system temperatures 
$k_BT=0.5$ and 2.2692($T_c$) calculating the deformation error (SSD error) 
of the bond energy along the rows as a function of $k_BT_{\rm eff}$. 
For $k_BT=0.5$, $f_{\rm pw}$ gives smaller deformation error than the other three, 
but for $k_BT=2.2692$ the same $f_{\rm pw}$ gives the largest SSD error. 
The ones for $f_{\rm sin}$ and $f_{\rm linear}$ also have different tendencies depending on $k_BT$. 
This is because, the deformation error overall tends to increase as the slope $d(k_BT_{\rm eff})/dr$ becomes larger 
as we discussed previously. 
The temperature slope at each $k_BT_{\rm eff}$ varies depending on the system temperature and 
the choice of envelope functions. 
However, since the slope of SSD varies with the moderate tendency compared to the other three cases, 
and having the optimal zero-slope $d(k_BT_{\rm eff})/dr=0$ at both $r/R=0$ and 1, 
it sustains as a moderately stable function, not depending much on the system parameters. 
Therefore, although one may choose other functions at their purposes, 
the SSD may be regarded as an optimal function in the sense that it does not require tuning of parameters. 
%
%
%
\section{Summary and Discussion} 
\label{sec_final}
We analyzed the 1D and 2D ferromagnetic Ising model with spatially deformed interactions 
in the sine-square functional form. 
We found that this interaction-deformed system is equivalent to the uniform-interaction system 
with the spatially deformed temperature. 
To be more precise, 
we showed by the analytical and numerical analyses 
that this deformed classical system is an assemblage of small subsystems. 
Each subsystem locally realizes the equilibrium of a uniform system at its own effective temperature. 
We propose that this classical SSD at finite temperature gives 
the approximate extended canonical ensemble with its ``state indices" 
spanning over real space. 
\par
In the analytical calculation, 
we first extended the formulation of the conventional transfer matrix method to those of the nonuniform system. 
We showed that the partition function 
is exactly obtained even though the interaction is deformed in one direction; 
this fact is rather trivial in 1D, as the eigenstate of the transfer matrix defined on each bond 
does not depend on the strength of the interactions. 
In 2D, the transfer matrices are defined in a unit of a column of the lattice of length $L$, 
describing the contributions from $2^L$ different configurations of the Ising variables. 
Referring to the previously established approach, 
the Ising variables are transformed to the noninteracting 1D fermionic operators, 
and $2^L$ ensemble average of the Ising variables are mapped to 
the summation of 2$^L$ different many-body states constructed from 
the noninteracting one-body states of Bogoliubov fermions. 
For demonstration, these formulas are numerically evaluated in 1D and 2D Ising models with SSD for system size 
$L=10\sim 500$. 
Notice that it is practically possible to extend it to $L\gtrsim 10000$ if needed, as it is a one-body problem of fermions on a chain of length $L$. 
In the uniform 2D system, the above mentioned Bogoliubov quasi-particles are 
itinerant plane waves characterized by wavenumbers, 
but once the SSD turns on, they mix via scattering of the SSD potential 
and form a set of spatially localized wave packet states. 
At the same time, according to our picture, 
{\it ``a system with spatially nonuniform interaction bonds $J_i$ 
at the temperature $k_BT$" }
could be interpreted as {\it ``a spatially uniform system with interaction bonds $J$ exposed to 
the spatially varying effective temperature 
$k_BT_{\rm eff}/J=k_BT/J_i$"}. 
Then, the quasi-particle localized on a certain bond feels  
the effective temperature, and carries the energy corresponding to 
that of the bulk system at $k_BT_{\rm eff}$. 
The trace of the product of the exponentials of these quasi-particle energy gives the partition function. 
The constituent of this product in a unit of the localized wave packet gives 
the local partition function, carrying the energy typical to that location. 
The system thus becomes a canonical ensemble of wave-packet states representing 
the thermal equilibrium at temperature $k_BT_{\rm eff}$. 
We showed that this picture is valid by evaluating the bond energy 
numerically exactly for a finite size $L$. 
Its deviation from the exact value in the thermodynamic limit is suppressed to less than $10^{-3}$. 
\par
A practical advantage of using the SSD is that it generates a massive number ($L/2$) 
of data points with different $k_BT_{\rm eff}$ by a single calculation at fixed $k_BT$. 
Therefore, we can perform Monte Carlo simulations on a large system, 
and obtain a wide profile of the energy and the specific heat within a 
sufficient accuracy with a low numerical cost. 
The SSD error is found to scale roughly with $1/\sqrt{L}$.
When combined with the Gaussian kernel regression,
the accuracy improves beyond the SSD error.
The applications of the SSD to the MC simulations are very promising. 

\section{acknowledgement}
We thank Hosho Katsura, Kenichi Asano, and Koji Hukushima for useful information. 
This work is supported by JSPS KAKENHI Grants (No. JP17K05533, No. JP18H01173, 
No.JP21K03440 and No. 20K03773) from the Ministry of Education,  Science, Sports and Culture of Japan.



\begin{references}
\bibitem{footnote2} One could also describe it in the Majorana fermions while in that case, 
the representation of the transfer matrix is given by an antisymmetric matrix which may be numerically rather bothersome to treat.
\end{references}
\end{document}